\documentclass[fleqn,usenatbib]{mnras}

\usepackage{newtxtext,newtxmath,nicefrac}

\usepackage[T1]{fontenc}

\DeclareRobustCommand{\VAN}[3]{#2}
\let\VANthebibliography\thebibliography
\def\thebibliography{\DeclareRobustCommand{\VAN}[3]{##3}\VANthebibliography}

\usepackage{graphicx}	
\usepackage{amsmath}	

\title[EBLM XIV.]{The EBLM project -- XIV. TESS light curves for eclipsing binaries with very low mass companions}
\author[J. Fitzpatrick et al.]
{Jay Fitzpatrick,$^{1,2\,\href{https://orcid.org/0009-0001-5149-4923}{\includegraphics[scale=0.5]{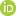}}}$
Pierre F. L. Maxted$^{1\,\href{https://orcid.org/0000-0003-3794-1317}{\includegraphics[scale=0.5]{orcid.jpg}}}$,\thanks{E-mail: p.maxted@keele.ac.uk}
Alix V. Freckelton$^{3\,\href{https://orcid.org/0009-0007-1053-0004}{\includegraphics[scale=0.5]{orcid.jpg}}}$,
A. H. M. J. Triaud$^{3\,\href{https://orcid.org/0000-0002-5510-8751}{\includegraphics[scale=0.5]{orcid.jpg}}}$,\newauthor
D.~V. Martin$^{4,5\,\href{https://orcid.org/0000-0002-7595-6360}{\includegraphics[scale=0.5]{orcid.jpg}}}$,
and 
A. Duck$^{5\,\href{https://orcid.org/0000-0002-4531-6899}{\includegraphics[scale=0.5]{orcid.jpg}}}$\\
$^{1}$Astrophysics Group, Keele University, Keele, Newcastle, ST5 5BG, UK\\
$^{2}$Department of Physics and Astronomy, University of Exeter, Exeter EX4 4QL, UK\\
$^{3}$School of Physics \& Astronomy, University of Birmingham, Edgbaston, Birmingham B15 2TT, UK\\
$^{4}$Department of Physics \& Astronomy, Tufts University, Medford, MA 02155, USA\\
$^{5}$Department of Astronomy, The Ohio State University, Columbus, OH 43210, USA\\
}

\date{Accepted XXX. Received YYY; in original form ZZZ}

\pubyear{2024}

\begin{document}
\label{firstpage}
\pagerange{\pageref{firstpage}--\pageref{lastpage}}
\maketitle

\begin{abstract}
Accurate limb-darkening models are needed for accurate characterisation of eclipsing binary stars and transiting exoplanets from the analysis of their light curves.
The limb-darkening observed in solar-type stars from the analysis of light
curves for transiting hot-Jupiter exoplanets are systematically less steep than
predicted by stellar model atmospheres that do not account for the stellar
magnetic field.
Hot-Jupiter host stars tend to be metal rich ([Fe/H] $\approx$
0.25) leading to a lack of low- and solar-metallicity targets in previous studies, so we have analysed the {\it TESS} light curves for a sample of 19 stars with transiting M-dwarf companions to extend the range of limb-darkening
measurements to [Fe/H] values more typical for solar-type stars.
We find that the systematic offset between the observed and predicted
limb-darkening profiles observed in metal-rich hot-Jupiter systems is also
observed for these solar-type stars at lower metallicity.
These observations provide additional measurements to explore the impact of
magnetic fields on the atmospheres of solar-type stars.
We have also used the {\it TESS} light curves to make precise estimates of the
radius and effective temperature of the M-dwarf companions in these 19 binary
systems.
We confirm the results from previous studies that find very low mass stars
tend to be about 3 per cent larger than predicted by stellar models that use a
mixing length prescription calibrated on the Sun.
\end{abstract}




\section{Introduction}

The availability of very high precision photometry from space mission such as  {\it Kepler} \citep{2010Sci...327..977B} and the Transiting Exoplanet Survey Satellite  \citep[{\it TESS},][]{2015JATIS...1a4003R} now make it possible to directly measure the limb-darkening of solar-type stars by analysing the light curves of transiting exoplanets.
The works particularly well for large planets on short-period orbits that produce large numbers of deep transits in the observed light curve, i.e. ``hot Jupiter'' systems.  
Previous studies have used {\it Kepler} and {\it TESS} light curves of hot-Jupiter systems to study the accuracy of limb-darkening profiles computed with stellar model atmospheres \citep{2007ApJ...655..564K,2013A&A...560A.112M, 2015MNRAS.450.1879E,2018A&A...616A..39M, 2022AJ....163..228P}. 
These studies have been motivated by the need to deal with limb-darkening in order to obtain accurate planet properties, so the coefficients of the limb-darkening laws used in this context are seen as ``nuisance parameters''.
However, limb-darkening is a direct probe of the temperature structure in a star's atmosphere, so there is scope to use limb-darkening measurements to study and improve stellar model atmospheres. 
It is not straightforward to interpret the coefficients of a limb-darkening law derived from a fit to a transit light curve, i.e. to understand which features of the star's limb-darkening  profile are constrained by the transit light curve. 
\cite{2023MNRAS.519.3723M} studied the limb-darkening information contained in the light curves of transiting exoplanets using a least-square fits of a transit model with a 4-parameter limb-darkening law \citep{2000A&A...363.1081C}. 
These least-squares fits were used to derive the following quantities for 43 solar-type stars:
\begin{equation*}
\begin{array}{ll}
h^{\prime}_1 = & I_{\lambda}(2/3),\\
\noalign{\smallskip}
h^{\prime}_2 = & h^{\prime}_1 - I_{\lambda}(1/3).
\end{array}
\end{equation*} 
These parameters were chosen because they are weakly correlated with one another, and they are directly related to  $I_\lambda(\mu)$, the specific intensity at some wavelength $\lambda$, as a function of $\mu = \cos(\theta)$, where $\theta$ is the angle between the line of sight and the surface normal vector.\footnote{For a spherical star, $\mu = \sqrt{1-r^2}$, where $r$ is the radial coordinate on the stellar disc from $r=0$ at the centre to $r=1$ at the limb.}
This makes it straightforward to compare the constraints on the limb-darkening profile from the observations to the predictions made by model stellar atmospheres.
These observations have subsequently been used to estimate the mean magnetic field strength in the atmospheres of these stars \citep{2023A&A...679A..65L, 2024NatAs.tmp...73K}.
In this study, we have increased the sample of solar-type stars with accurate limb-darkening measurements by applying the method devised by \cite{2023MNRAS.519.3723M} to the {\it TESS} light curves of 19 eclipsing binaries with low-mass companions. 
These ``EBLM'' systems have similar light curves to hot-Jupiter systems because the radius of a star at the bottom of the main-sequence is comparable to a strongly-irradiated gas-giant planet in a short-period orbit around a solar-type star \citep{2023Univ....9..498M}.
This allows us to use the same assumptions for EBLMs as we do for hot-Jupiters. 
Many detailed studies into observed limb-darkening focus on using transiting hot-Jupiters and the biases introduced by selecting different laws on the parameters of the planet and system (e.g. \citet{2016MNRAS.457.3573E}). Previous studies looking into discrepancies between theoretical limb-darkening parameters and observed profiles are currently biased to host stars with high metallicity \citep{2023MNRAS.519.3723M}.
The hot-Jupiter systems studied in \citet{2023MNRAS.519.3723M} tend to be metal rich ($[{\rm Fe/H}] \approx 0.25$), leading to their being only two {\it TESS} targets below solar metallicity in that sample. 
This bias is not present for the stars in EBLM systems, so this new sample provides data on the limb-darkening properties of stars over a wider range of metallicity. 
\citet{2024MNRAS.534.3893V} have recently re-analysed the same targets studied by \cite{2023MNRAS.519.3723M} using a regularisation technique that improves the precision in the measured limb-darkening by a factor of up to 2. We have not attempted to use this regularisation technique for our analysis because the precision of the results we have obtained is sufficient for us to achieve our aim to test whether metallicity has an impact on the accuracy of the limb-darkening estimates from models compared to observations.

The data available for the EBLMs in this study also allow us
to infer the mass, radius and effective temperature of the low-mass stars in these binary systems.
These are valuable fundamental data that can be used to assess the accuracy of stellar evolution models for very low mass stars, e.g. to investigate the long-standing ``radius inflation'' problem -- the tendency for real stars to be larger and cooler than predicted by stellar models when the mixing-length  parameter used to model convection is calibrated to the solar value \citep{1970ApJ...161.1083H, 1997AJ....114.1195P,2013ApJ...776...87S}.
The leading hypothesis explaining the apparent inflation of M-dwarf radii is strong magnetic fields altering the transport of convective energy in the interior of low-mass stars \citep{2001ApJ...559..353M,2007A&A...472L..17C,2014ApJ...787...70M,2013ApJ...779..183F,2014ApJ...789...53F}.

\section{Methods}

\subsection{Target selection} \label{subsec:targets}

\begin{figure}
	\includegraphics[width=\columnwidth]{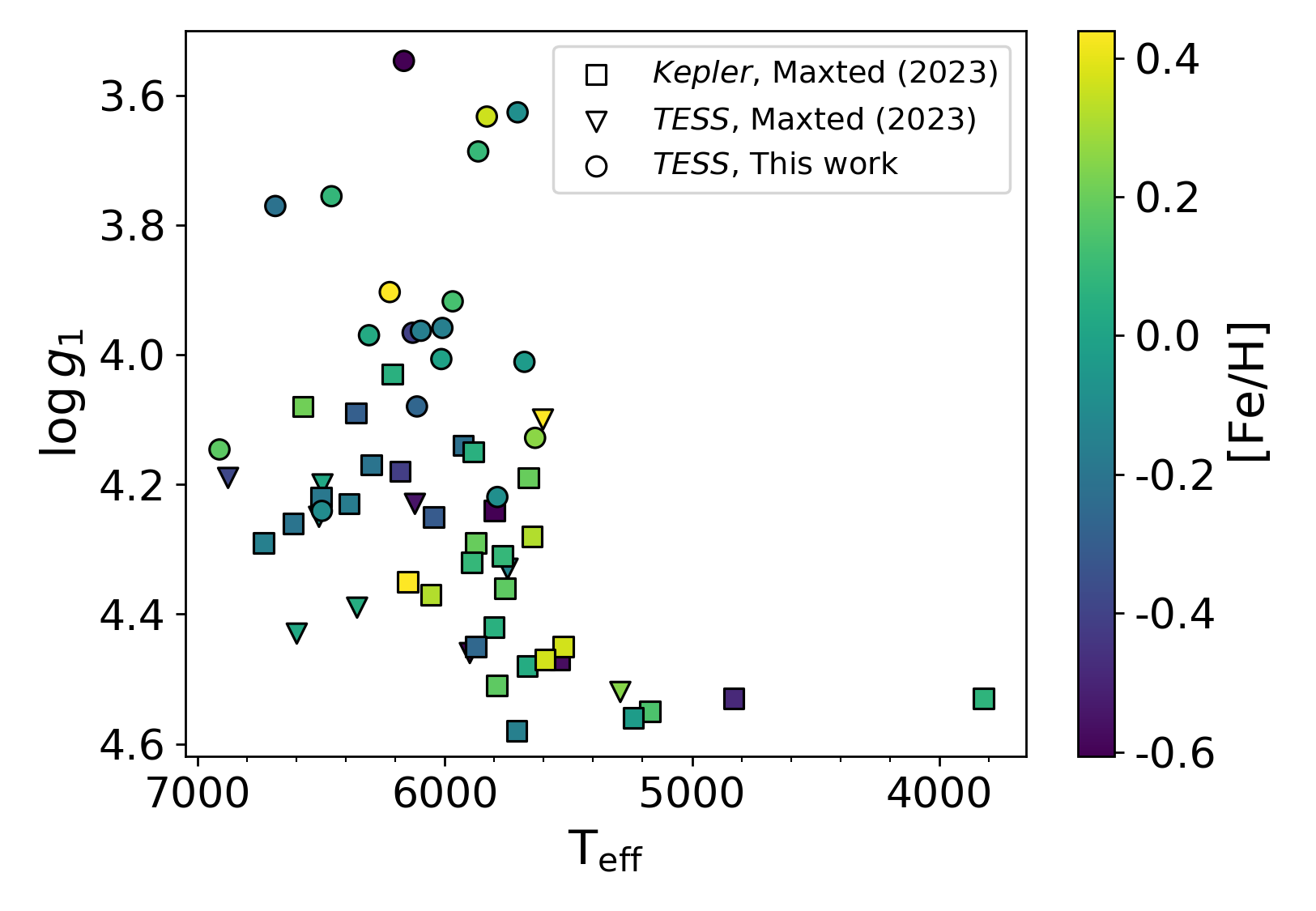} 
    \caption{Kiel diagram of EBLMs and stars with transiting hot-Jupiters investigated in this paper. Triangle and circle points give targets observed with \textit{TESS} from \citet{2023MNRAS.519.3723M} and this work respectively. Square points give targets investigated in the \textit{Kepler} band in \citet{2023MNRAS.519.3723M}.} 
    \label{fig:kiel}
\end{figure}

The initial list of EBLM targets were taken from objects of interest found in the BEBOP survey \citep{2019A&A...624A..68M} and the EBLM project \citep{2017A&A...608A.129T}. A light curve covering many primary eclipses are needed to achieve the required signal-to-noise ratio for an accurate measurement of the limb-darkening with {\it TESS} data, so we selected targets with at least four sectors of $120$-s cadence {\it TESS} data available and orbital periods $P<25$\,d. Multiple visits of the secondary eclipses of a target are also important to determine the depth as secondary eclipses of EBLMs are on the order of $0.1\%$ of the total flux of the system.
Light curves were visually inspected using {\sc lightkurve} \citep{2018ascl.soft12013L}.\footnote{\url{https://docs.lightkurve.org/}}
Systems with evidence for strong magnetic activity from pseudo-sinusoidal variations in the light curve  were omitted.
It is difficult to characterise the limb-darkening systems with a  high impact parameter \citep[$b \gtrsim 0.8$, ][]{2013A&A...560A.112M} so  
these systems were also excluded via a literature search. 
Where published impact parameters were not available, primary and secondary eclipses were visually inspected. 
Systems with primary eclipses that were not well defined or had no clear detection of a secondary eclipse were also excluded. 
The final list of 19 targets can be seen in Table~\ref{tab:params}, including the {\it TESS} input catalogue (TIC) number \citep{2019AJ....158..138S}.
The sample of stars from \citet{2023MNRAS.519.3723M} with limb-darkening measurements in the {\it TESS} band have a metallicity range of [Fe/H]\,$=-0.11$ to +0.44 with an average of [Fe/H]\,$ = +0.16$. This sample of EBLM targets have a range of [Fe/H]\,$=-0.61$ to +0.32 with an average of [Fe/H]\,$=-0.097$ (see section~\ref{sec:teff}). An overview of all targets investigated in this paper can be seen in Fig~\ref{fig:kiel}.

\begin{table}
\centering
\caption{EBLM identifiers and TIC numbers, V-band magnitudes, orbital periods ($P$) and and time of mid primary eclipse ($T_{0,{\rm pri}}$) for our sample of eclipsing binary stars..}
 \label{tab:params}
\begin{tabular}{@{}lrrrrr}
\hline		
EBLM  &
  \multicolumn{1}{l}{TIC} &
  \multicolumn{1}{c}{$V$} & 
  \multicolumn{1}{c}{$P$} &
  \multicolumn{1}{c}{BJD$_{\rm TDB}$ $T_{0,{\rm pri}}$} \\ 
  && [mag.]& [days] & [$-2457000$]\\
		\hline
J0228$+$05	& 422844353	& 10.23	& 6.634710	& 2162.9272	\\
J0247$-$51	& 231275247	& 9.55	& 4.007899	& 2087.5593	\\
J0400$-$51	& 237342298	& 12.27	& 2.692083	& 2121.2505	\\
J0432$-$33	& 170749770	& 11.14	& 5.305547	& 1460.2140	\\
J0440$-$48	& 259470701	& 11.52	& 2.543001	& 2144.8976	\\
J0500$-$46	& 161577376	& 12.03	& 8.284416	& 1487.5124	\\
J0526$-$34	& 24397947	& 11.20	& 10.190907	& 1489.7715	\\
J0608$-$59	& 260128333	& 11.98	& 14.608556	& 2140.2922	\\
J0625$-$43	& 232030067	& 12.11	& 3.969008	& 2199.2756	\\
J0627$-$67	& 167201539	& 11.52	& 9.468921	& 2228.0884	\\
J0709$-$52	& 343963409	& 13.18	& 9.108032	& 2214.5151	\\
J0723$+$79	& 289949453	& 9.26	& 13.217665	& 2590.4378	\\
J0829$+$66	& 102841799	& 12.03	& 5.260395	& 2583.7945	\\
J0941$-$31	& 25776767	& 11.08	& 5.545649	& 2262.3175	\\
J0955$-$39	& 45599777	& 12.87	& 5.313593	& 2264.1666	\\
J1626$+$57	& 207493152	& 10.61	& 9.127609	& 2644.7358	\\
J1640$+$49	& 274238240	& 10.38	& 6.206219	& 2668.8626	\\
J1705$+$55	& 198358825	& 12.66	& 23.514357	& 2576.5598	\\
J1850$+$50	& 48191564	& 12.21	& 9.569037	& 2416.6780	\\
 \hline
	\end{tabular}
\end{table}

\subsection{Pre-processing} \label{subsec:preprocessing}
{\sc lightkurve} was used to download \textit{TESS} pre-search data conditioning simple aperture photometry (PDCSAP) with a cadence of 120\,s for each star.  
Data points outside 10 standard deviations ($>10\,\sigma$) from the median of the data set were  excluded. 
Data within 1.5 times of the respective eclipse widths relative to the time of mid eclipse were kept for fitting.
The observation time of the light curves were folded at the period to give data points in phase units with phase zero equal to the time of mid primary eclipse. 
For the folded primary and secondary eclipses, the data were binned in 120-s intervals using the median value flux in each phase bin with errors equal to the 1.25 times the mean of the absolute deviation from the median of fluxes in the same phase bin.
Unbinned data points that fell outside of 5 times the standard error on their bin were removed.

A straight line was fit to the data either side of each primary and secondary eclipse and used to flatten the light curves by dividing through the data, bringing all the points onto a consistent flux scale. These straight line fits were visually inspected before proceeding to check the straight line fit to the flux outside of eclipse. This also allowed us to check eclipses were properly aligned around the respective times of mid eclipse, ensuring there was no misalignment when combining all the observations into one light curve containing the primary eclipses and one containing the secondary eclipses as a function of phase.
We also checked for and removed data from orbital cycles containing bad data, e.g. poor correction of instrumental trends that sometimes occurs near gaps in the data. This results in light curves like the grey data points on Fig~\ref{fig:lcfit}.

\subsection{Generating the light curve model}
We use {\sc batman} version 2.4.9 by \citep{2015PASP..127.1161K} to model the primary and secondary eclipses in the observed light curves assuming Claret's 4-parameter non-linear law to parameterise the limb-darkening profile for the primary star \citep[][]{2000A&A...363.1081C}:
\begin{equation}
    \label{eqn:claret}
    I_{\lambda}(\mu) = 1 - \sum_{j=1}^{4}a_j(1-\mu^{j/2}).
\end{equation}
This limb-darkening law was chosen as it provides the sufficient flexibility to capture the true limb-darkening profile, without making strong assumptions about the shape compared to 2-parameter limb-darkening laws \citep{2016MNRAS.457.3573E,2023MNRAS.519.3723M}.
This allows the parts of the limb-darkening profile that are weakly constrained by the light curve to have large uncertainties.

{\sc batman} computes primary and secondary eclipses separately. We combined these models to simultaneously fit both features in the light curve. The  primary eclipse model from {\sc batman}  is normalised and only takes into account the flux from the primary star. 
The secondary eclipse model sets the primary flux level to one with the flux from the companion added, as set by the flux ratio parameter. 
The secondary eclipse model does not take into account for R{\o}mer delay, which arises from the finite speed of light causing a delay on the order of a few seconds when the secondary star is eclipsed by the primary star. We used the following equation 
\citep{2010arXiv1006.3834F,2015MNRAS.448..946B} to compute the R{\o}mer delay for the time of secondary eclipse due to light travel time across the orbit:
\begin{equation}
    \label{eqn:LTTecc}
    LTT = \frac{a \sin(i)}{c} \frac{1-q}{1+q} \frac{1-e^2}{1-e^2 \sin^2{\omega}}.
\end{equation}
Here, $a$ is the orbital semi-major axis, 
$i$ is the orbital inclination,
$c$ is the speed of light, 
$q=M_2/M_1$ is the mass ratio, 
$e$ is the eccentricity and $\omega$ is the argument of periastron.

The following steps were then used to calculate the model.
\begin{enumerate}
    \item Primary and secondary eclipse light curve models calculated.
    \item Total primary star flux (=1) taken away from the secondary eclipse light curve model, leaving us with just the flux from the secondary star.
    \item Primary and secondary eclipse light curve models added together.
    \item Total light curve re-normalised so the flux outside of eclipse is equal to one (i.e. primary flux plus secondary flux is equal to one)
\end{enumerate}

\subsection{Primary and secondary eclipse fitting} \label{subsec:fit}
The free parameters in the fit were: orbital period, $P$; time of mid-primary eclipse, $T_{0,{\rm pri}}$; secondary-primary star radius ratio, $k=R_{2}/R_{1}$; fractional radius of the primary star $R_{1}/a$; the impact parameter,\footnote{$b$ is not exactly equal to the true impact parameter for eccentric orbits because this depends on $e$ and $\omega$ and has different values for the transit and eclipse \citep{2010exop.book...55W}. This complication has been ignored for this study. } $b = a\cos{i}/R_{1}$;
flux ratio in the {\it TESS} band, $L_T = f_{\rm sec}/f_{\rm pri}$; the coefficients $a_1,\dots,a_4$ of the Claret 4-parameter limb-darkening law  (equation~(\ref{eqn:claret})). 
For systems with eccentric orbits, $e$ and $\omega$ or $e\cos(\omega)$ and $e\sin(\omega)$ were also varied depending on whether prior measurements of these quantities are available. For stars with measured $e$ and $\omega$, these parameters were varied with Gaussian priors to remain consistent with the measured values. For stars without these measurements, we used $e\cos(\omega)$ and $e\sin(\omega)$ as free parameters because these quantities are less correlated with each other than $e$ and $\omega$.
Following the methods in \citet{2013PASP..125...83E}, this requires a ``change of variables'' transformation for the calculation of the joint prior probability distribution function for $e\cos(\omega)$ and $e\sin(\omega)$ to ensure the implied prior on $e$ is uniform.
This was achieved using the following equation,  where $P(e\cos(\omega),e\sin(\omega))$ is the prior on $e\cos(\omega)$ and $e\sin(\omega)$ and $P(e,\omega)$ is the prior on $e$ and $\omega$:
\begin{equation}
    \label{eqn:transform}
    P(e\cos(\omega),e\sin(\omega)) = \frac{P(e,\omega)}{e}.
\end{equation}

To sample the posterior probability distribution (PPD) for the vector of model parameters $\bmath{\theta}$ given the data (light curve), $\cal{D}$, $P(\bmath{\theta} | \cal{D}) \propto P(\cal{D}|\bmath{\theta})P(\bmath{\theta})$ we used the affine-invariant Markov chain Monte Carlo sampler {\sc emcee} \citep{2013PASP..125..306F, 2010CAMCS...5...65G}. 
To compute the likelihood $P(\cal{D}|\bmath{\theta})$ we assume that the error on data point $i$ has a Gaussian distribution with standard deviation $f\sigma_i$, and that these errors are independent. 
The logarithm of the error-scaling factor $f$ is included as a hyper-parameter in the vector of model parameters $\bmath{\theta}$. 
We used uniform priors on all model parameters (excluding special cases for $e\cos(\omega)$ and $e\sin(\omega)$ discussed above) within the full range allowed by the {\sc batman} model.

\begin{figure}
	\includegraphics[width=\columnwidth]{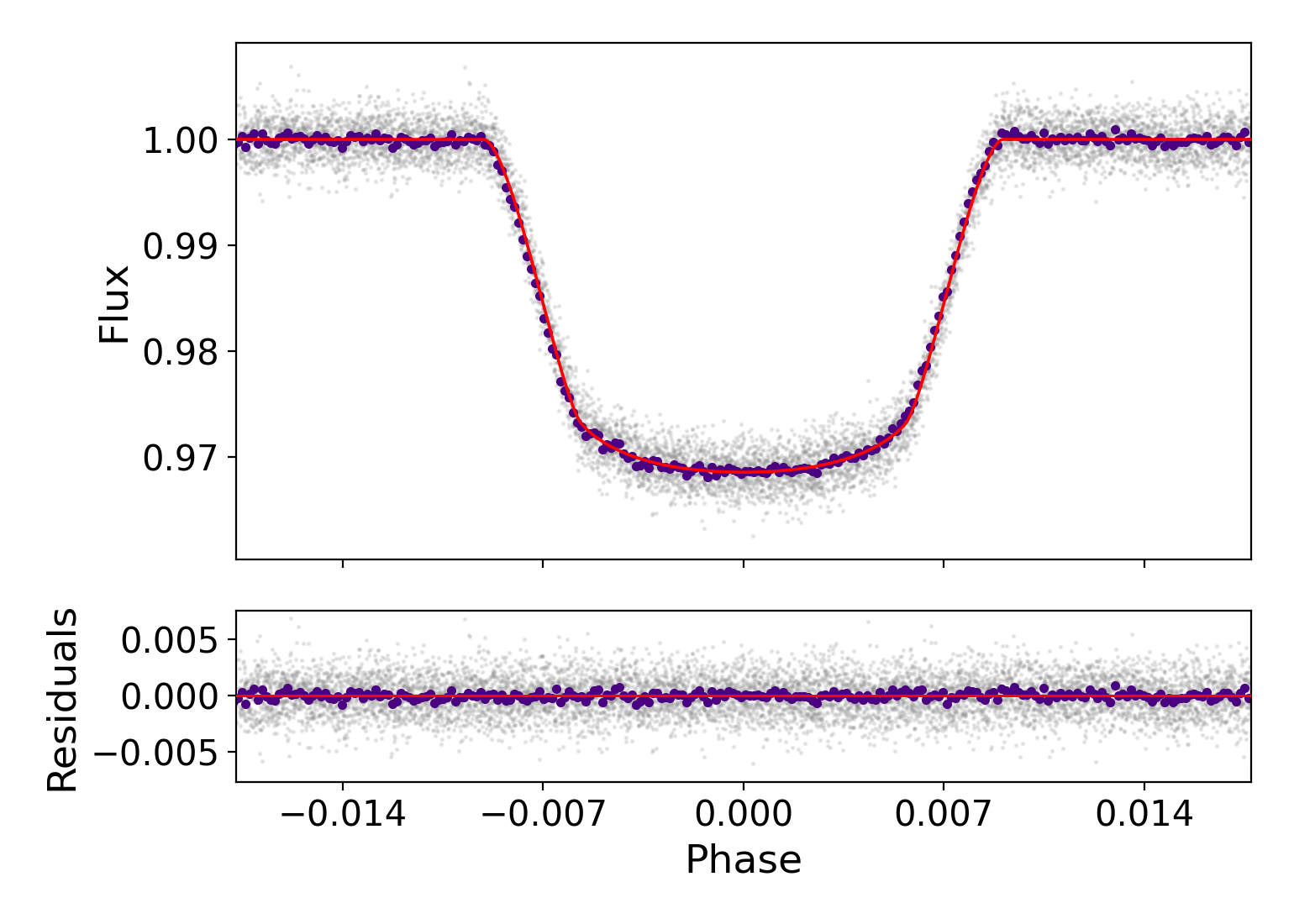}
	\includegraphics[width=\columnwidth]{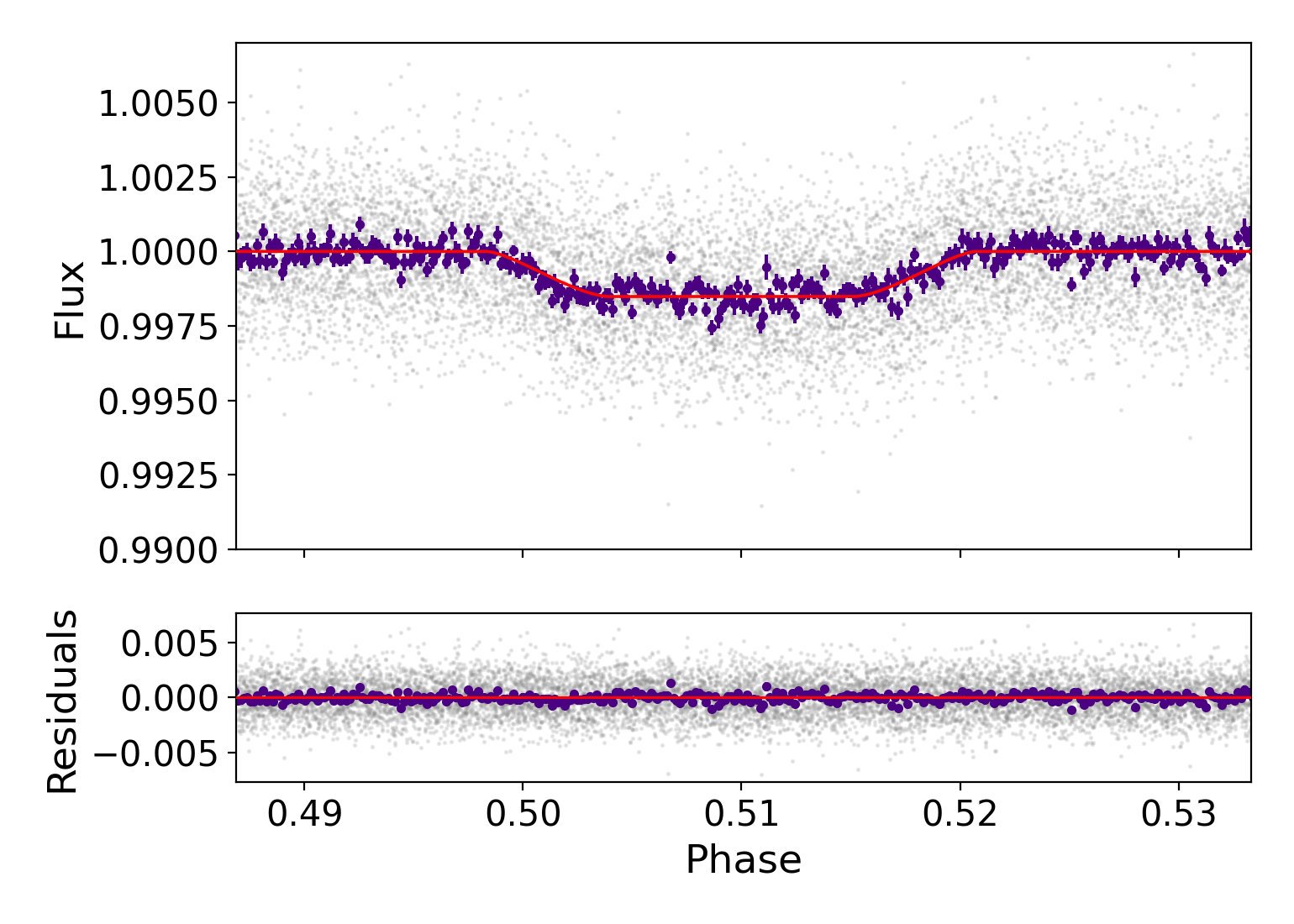}
    \caption{The {\it TESS} light curve of EBLM~J0627$-$67 with the best fit from our analysis (red line) to unbinned data points (grey). The two top panels show the fit and residuals from the fit to the primary eclipse, and the bottom two panels show show the fit and residuals from the fit to the secondary eclipse. Purple data points show the light curve data binned in 120\,s intervals for reference.} 
    \label{fig:lcfit}
\end{figure}

We used 100 walkers and 1000 steps to generate a random sample of points from the PPD following a minimum of 4000 ``burn-in'' steps. 
Convergence of the chain was confirmed by visual inspection of the sample values for each parameter as a function of step number to ensure that there are no trends in the mean values or variances for the sample values from all walkers after the burn-in phase.
This was verified using the {\tt get\_autocorr\_time} function in {\sc emcee} and a number of burn in steps equivalent to at least a few times the autocorrelation time for the parameters was used.
4\,000 steps was sufficient to achieve convergence for all systems apart from EBLM~J0432$-$33, for which we used 19\,000 burn-in steps. 
The slow convergence for this system is due to $b$ having a bimodal PPD for $b$ with peaks near $b=\pm 0.2$.
While bimodal distributions in MCMC analysis can cause problems, as a star transiting at $b = -0.1,-0.2,-0.3\dots$ produces the same light curve as a star transiting at $b = 0.1,0.2,0.3\dots$ respectively (assuming radial symmetry), the absolute value of the chains can be taken to get summary statistics. This was done for EBLM~J0400$-$51, EBLM~J0432$-$33, EBLM~J0608$-$59 and EBLM~J1640$+$49. Summary statistics are given in Table~\ref{tab:fitresults} where parameter values are the median of the sample and the errors are based on the 15.87\% and 84.13\% percentiles of the sample (one standard deviation of a normal distribution). The fit for EBLM~J0627$-$67 is given as an example in Fig~\ref{fig:lcfit}.

\begin{table*}
\centering
\caption{Parameters obtained from {\sc emcee} fits of the {\sc batman} model to {\it TESS} light curves of EBLM binaries. Values in parentheses are the standard errors on the final two digits of the preceding value, with the parameters taken as the median of the {\sc emcee} chain and the errors based on the 15.87\% and 84.13\% percentiles. $J_T = L_T/k^2$ is the surface brightness ratio in the {\it TESS} band.
}
 \label{tab:fitresults}
\begin{tabular}{@{}lrrrrrrrrrrr}
\hline		
EBLM  & 
  \multicolumn{1}{c}{$h^{\prime}_1$ } & 
  \multicolumn{1}{c}{$h^{\prime}_2$ } &
  \multicolumn{1}{c}{$D$ } & 
  \multicolumn{1}{c}{$W$} &
  \multicolumn{1}{c}{$b$} &
  \multicolumn{1}{c}{$L_T$ } & 
  \multicolumn{1}{c}{$k=R_2/R_1$} &
  \multicolumn{1}{c}{$R_1/a$} & 
  \multicolumn{1}{c}{$e$} & 
  \multicolumn{1}{c}{$J_T$} \\ 
		\hline
 J0228$+$05 & $   0.901 (9)$ & $    0.13 (2)$ &	$   0.0165 (1)$&	$  0.02985 (9)$&	$     0.41 (2)$&	$  0.00063 (3)$&	$   0.1284 (5)$&	$   0.0893 (7)$&	      $-$   	&	$    0.038 (2)$\\
 J0247$-$51 & $   0.903 (3)$ & $    0.13 (1)$ &	$   0.0189 (1)$&	$   0.0518 (1)$&	$     0.27 (2)$&	$  0.00111 (1)$&	$   0.1376 (4)$&	$   0.1474 (7)$&	$    0.004 (1)$&	$   0.0586 (8)$\\
 J0400$-$51 & $   0.890 (5)$ & $    0.12 (2)$ &	$   0.0275 (2)$&	$   0.0662 (2)$&	$     0.06 (5)$&	$  0.00171 (5)$&	$   0.1658 (5)$&	$   0.1787 (8)$&	      $-$   	&	$    0.062 (2)$\\
 J0432$-$33 & $   0.882 (5)$ & $    0.15 (2)$ &	$   0.0147 (2)$&	$   0.0511 (2)$&	$     0.17 (8)$&	$  0.00093 (3)$&	$   0.1213 (7)$&	$    0.145 (1)$&	      $-$   	&	$    0.063 (2)$\\
 J0440$-$48 & $    0.93 (1)$ & $    0.17 (3)$ &	$   0.0159 (2)$&	$   0.0634 (3)$&	$     0.49 (2)$&	$  0.00092 (3)$&	$   0.1260 (8)$&	$    0.196 (2)$&	      $-$   	&	$    0.058 (2)$\\
 J0500$-$46 & $    0.88 (3)$ & $    0.15 (3)$ &	$   0.0218 (3)$&	$   0.0228 (1)$&	$     0.62 (1)$&	$  0.00121 (5)$&	$    0.148 (1)$&	$   0.0740 (8)$&	$   0.2298 (4)$\makebox[0pt][l]{$^{\rm a}$}&	$    0.056 (2)$\\
 J0526$-$34 & $   0.880 (4)$ & $    0.19 (3)$ &	$   0.0272 (3)$&	$   0.0306 (1)$&	$     0.24 (3)$&	$  0.00214 (4)$&	$    0.165 (1)$&	$   0.0844 (5)$&	$   0.1261 (2)$\makebox[0pt][l]{$^{\rm a}$}&	$    0.079 (1)$\\
 J0608$-$59 & $   0.880 (2)$ & $    0.16 (2)$ &	$   0.0544 (3)$&	$  0.01827 (4)$&	$     0.09 (5)$&	$  0.00411 (3)$&	$   0.2332 (7)$&	$   0.0467 (2)$&	$   0.1562 (3)$\makebox[0pt][l]{$^{\rm a}$}&	$   0.0756 (6)$\\
 J0625$-$43 & $   0.885 (4)$ & $    0.15 (2)$ &	$   0.0299 (3)$&	$   0.0547 (2)$&	$     0.19 (3)$&	$  0.00248 (4)$&	$   0.1730 (9)$&	$   0.1484 (9)$&	      $-$   	&	$    0.083 (1)$\\
 J0627$-$67 & $    0.90 (1)$ & $    0.19 (2)$ &	$   0.0285 (2)$&	$  0.02029 (9)$&	$    0.583 (9)$&	$  0.00151 (3)$&	$   0.1687 (6)$&	$   0.0629 (3)$&	$   0.1587 (6)$\makebox[0pt][l]{$^{\rm a}$}&	$   0.0529 (9)$\\
 J0709$-$52 & $    0.85 (6)$ & $    0.15 (4)$ &	$   0.0330 (7)$&	$   0.0323 (2)$&	$    0.583 (8)$&	$  0.00281 (8)$&	$    0.182 (2)$&	$   0.0986 (9)$&	$    0.343 (1)$\makebox[0pt][l]{$^{\rm a}$}&	$    0.085 (3)$\\
 J0723$+$79 & $   0.872 (4)$ & $    0.17 (1)$ &	$   0.0389 (2)$&	$  0.01916 (4)$&	$    0.405 (6)$&	$  0.00251 (1)$&	$   0.1974 (6)$&	$   0.0534 (1)$&	$    0.063 (3)$	&	$   0.0643 (5)$\\
 J0829$+$66 & $   0.879 (10)$ & $    0.16 (4)$ &	$   0.0503 (8)$&	$   0.0253 (2)$&	$     0.34 (2)$&	$  0.00352 (9)$&	$    0.224 (2)$&	$   0.0675 (6)$&	$    0.161 (5)$	&	$    0.070 (2)$\\
 J0941$-$31 & $   0.895 (9)$ & $    0.14 (2)$ &	$   0.0180 (2)$&	$   0.0410 (2)$&	$     0.38 (3)$&	$  0.00120 (3)$&	$   0.1340 (7)$&	$    0.121 (1)$&	$   0.1984 (4)$\makebox[0pt][l]{$^{\rm b}$}&	$    0.067 (2)$\\
 J0955$-$39 & $    0.89 (2)$ & $    0.10 (4)$ &	$   0.0460 (7)$&	$   0.0266 (1)$&	$     0.44 (3)$&	$   0.0029 (1)$&	$    0.214 (2)$&	$   0.0740 (8)$&	      $-$   	&	$    0.063 (3)$\\
 J1626$+$57 & $   0.879 (3)$ & $    0.16 (2)$ &	$   0.0215 (2)$&	$   0.0258 (1)$&	$     0.21 (3)$&	$  0.00088 (2)$&	$   0.1465 (6)$&	$   0.0719 (5)$&	$    0.039 (2)$	&	$   0.0409 (8)$\\
 J1640$+$49 & $   0.888 (4)$ & $    0.13 (1)$ &	$  0.01466 (6)$&	$   0.0414 (3)$&	$     0.04 (4)$&	$  0.00064 (2)$&	$   0.1211 (3)$&	$   0.1163 (9)$&	$    0.065 (8)$	&	$    0.044 (1)$\\
 J1705$+$55 & $   0.882 (5)$ & $    0.17 (2)$ &	$   0.0409 (4)$&	$  0.01431 (5)$&	$     0.32 (2)$&	$  0.00366 (4)$&	$   0.2023 (9)$&	$   0.0388 (2)$&	$   0.2646 (1)$\makebox[0pt][l]{$^{\rm c}$}&$    0.090 (1)$\\
 J1850$+$50 & $   0.867 (6)$ & $    0.15 (3)$ &	$   0.0252 (3)$&	$   0.0301 (3)$&	$     0.08 (6)$&	$  0.00147 (5)$&	$   0.1589 (9)$&	$    0.082 (1)$&	$    0.108 (8)$	&	$    0.058 (2)$\\
 \hline
 \multicolumn{8}{@{}l}{($^{\rm a}$) $e$ and $\omega$ priors from \citet{2017A&A...608A.129T}.}\\
 \multicolumn{8}{@{}l}{($^{\rm b}$) $e$ and $\omega$ priors from \citet{2024MNRAS.528.5703S}.}\\
 \multicolumn{8}{@{}l}{($^{\rm c}$) $e$ and $\omega$ priors from \citet{2022yCat.1357....0G}.}\\
\end{tabular}
\end{table*}

\subsection{Estimating the mean offset accounting for additional scatter} \label{sec:combine}
In the following discussion we frequently wish to measure an offset between estimates of parameters from two different sources. The python package {\sc pycheops}\footnote{https://pycheops.readthedocs.io/en/latest/index.html} contains  the {\tt combine} function which uses the following method appropriate for this type of calculation.

We assume that the measurements of the parameter in question (e.g. $h^{\prime}_1$,  $h^{\prime}_2$ or $R_2$) have some extra scatter beyond their quoted standard errors. This extra scatter, $\sigma_{\rm ext}$, will be a combination of variance of astrophysical origin, e.g. due to magnetic activity on the star, and systematic errors e.g. imperfect removal of instrumental noise. If we assume that all errors are independent and have a Gaussian distribution then the log-likelihood to obtain the observed differences $\bmath{\Delta } = \{\Delta_i \pm \sigma_i, i=1,\dots,N\}$ is 
 \[\ln\,p(\bmath{\Delta}\,|\,\langle\Delta\rangle,\sigma_{\rm ext}) = -\frac{1}{2} \sum_i \left[
    \frac{(\Delta_i-\langle\Delta\rangle)^2}{s_i^2}
    + \ln \left ( 2\pi\,s_i^2 \right )
\right], \]
where $s_i^2 = \sigma_i^2 + \sigma_{\rm ext}^2$. We assume a broad uniform prior on the mean offset, $\langle\Delta\rangle$, and a broad uniform  prior on $\ln \sigma_{\rm ext}$. We then sample the posterior probability distribution using {\sc emcee} for 128 walkers and a number of steps and ``burn-in'' steps stated in the relevant sections. We use the remaining sample to calculate the mean and standard deviation of the posterior probability distribution for $\langle\Delta\rangle$, i.e. the best estimate for the value of the offset and its standard error.

We note that a global offset may not always be applicable to quantify the difference between our observations and the models discussed in Section~\ref{subsec:ld_results}, as seen in Fig~\ref{fig:phoenixtrend}. 
This is investigated in Section~\ref{subsubsec:ld_teff}.

\subsection{Effective temperature and metallicity estimates}
\label{sec:teff}
Targets noted in Table \ref{tab:alixparam} as having parameters derived in this work were analysed using high-resolution spectra (R$\approx$60,000) observed between May 2008 and September 2014 for \citet{2017A&A...608A.129T} with the CORALIE spectrograph, mounted on the Swiss 1.2-m Euler telescope at La Silla, Chile \citep{Queloz2000}. All spectra were reduced using the CORALIE DRS version 3.4, and co-added to produce a single spectrum with high signal to noise ratio for each target. Following the method outlined by \citet{2024MNRAS.531.4085F}, initial estimates of spectroscopic parameters were determined using the curve-of-growth equivalent widths method, and subsequently final parameters were obtained through spectral synthesis. Both methods were implemented through the {\sc paws} pipeline\footnote{\url{https://github.com/alixviolet/PAWS}} using the iSpec package \citep{BlancoCuaresma2014, BlancoCuaresma2019}, with the {\sc spectrum} line list \citep{gray1994} and ATLAS model atmospheres \citep{kurucz2005}.

\section{Comparison to limb-darkening profiles from stellar atmospheric models} \label{sec:ld_comp}

\begin{table*}
    \centering
    \caption{Effective temperatures, gravities and metallicities with relevant references. Parameters are used for comparison in Section~\ref{sec:ld_comp} and to find secondary star parameters and investigate the radius inflation problem in Section~\ref{sec:rad_infl}.} 
    \label{tab:alixparam}
\begin{tabular}{lcccc}
\hline
EBLM  & 
  \multicolumn{1}{c}{$T_\text{eff}\,[\text{K}]$}&
  \multicolumn{1}{c}{$\log g_1$}&
  \multicolumn{1}{c}{$[\text{Fe/H}]$}&
  \multicolumn{1}{c}{Reference}\\
  \hline
J0228$+$05 & $6912\pm134$ & $4.15\pm0.14$ & $+0.08\pm0.08$ & \citet{2024MNRAS.531.4085F}\\
J0247$-$51 & $6130\pm191$ & $3.97\pm0.12$ & $-0.43\pm0.33$ & This study\\
J0400$-$51 & $6010\pm180$ & $3.96\pm0.10$ & $-0.21\pm0.19$ & This study\\
J0432$-$33 & $6166\pm134$ & $3.55\pm0.14$ & $-0.61\pm0.05$ & \citet{2024MNRAS.531.4085F}\\
J0440$-$48 & $6113\pm286$ & $4.08\pm0.13$ & $-0.31\pm0.37$ & This study\\
J0500$-$46 & $5788\pm116$ & $4.22\pm0.18$ & $-0.15\pm0.05$ & \citet{2024MNRAS.531.4085F}\\%
J0526$-$34 & $6307\pm128$ & $3.97\pm0.15$ & $-0.05\pm0.07$ & \citet{2024MNRAS.531.4085F}\\
J0608$-$59 & $5865\pm103$ & $3.69\pm0.25$ & $+0.01\pm0.05$ & \citet{2024MNRAS.531.4085F}\\
J0625$-$43 & $5678\pm118$ & $4.01\pm0.12$ & $-0.10\pm0.16$ & This study\\
J0627$-$67 & $6498\pm117$ & $4.24\pm0.27$ & $-0.16\pm0.13$ & \citet{2024MNRAS.531.4085F}\\
J0709$-$52 & $6097\pm312$ & $3.96\pm0.10$ & $-0.21\pm0.39$ & This study\\%
J0723$+$79 & $6014\pm121$ & $4.01\pm0.19$ & $-0.08\pm0.03$ & \citet{2024MNRAS.531.4085F}\\
J0829$+$66 & $5635\pm126$ & $4.13\pm0.15$ & $+0.16\pm0.06$ & \citet{2024MNRAS.531.4085F}\\
J0941$-$31 & $6686\pm134$ & $3.77\pm0.18$ & $-0.26\pm0.04$ & \citet{2024MNRAS.531.4085F}\\
J0955$-$39 & $6223\pm165$ & $3.90\pm0.66$ & $+0.32\pm0.22$ & This study\\
J1626$+$57 & $5968\pm118$ & $3.92\pm0.17$ & $+0.04\pm0.04$ & \citet{2024MNRAS.531.4085F}\\
J1640$+$49 & $6459\pm127$ & $3.76\pm0.17$ & $+0.01\pm0.04$ & \citet{2024MNRAS.531.4085F}\\
J1705$+$55 & $5706\pm112$ & $3.63\pm0.19$ & $-0.15\pm0.06$ & \citet{2024MNRAS.531.4085F}\\
J1850$+$50 & $5830\pm103$ & $3.63\pm0.16$ & $+0.24\pm0.04$ & \citet{2024MNRAS.531.4085F}\\
\hline
\end{tabular}
\end{table*}
Predicted values of $h^{\prime}_1$ and $h^{\prime}_2$ have been extracted for each target from published tabulations using linear interpolation to the values of $T_{\rm eff,1}$, $\log g_1$ and $[\text{Fe/H}]$ given in Table~\ref{tab:alixparam}. This was done for the following grids of model limb-darkening coefficients: ATLAS \citep{2017A&A...600A..30C}; \text{MPS-ATLAS} Set 1 \citep{2022A&A...666A..60K}; \text{MPS-ATLAS} Set 2 \citep{2022A&A...666A..60K}; \text{PHOENIX-COND} \citep{2018A&A...618A..20C} and \text{Stagger-grid} \citep{2018A&A...616A..39M}. Errors were determined using a Monte Carlo method, assuming Gaussian independent errors on input values. The grids for \citet{2017A&A...600A..30C}, \citet{2018A&A...618A..20C} and \citet{2018A&A...616A..39M} give the limb-darkening coefficients, $a_1,\dots,a_4$. 
These were extracted using a linear interpolation and then equation~(\ref{eqn:claret}) was used to give the specific intensities so that we could calculate $h^{\prime}_1$ and $h^{\prime}_2$.

The limb-darkening profiles for \text{PHOENIX-COND} models from \citet{2018A&A...618A..20C} are calculated on a grid that extends beyond the limb of the star so the limb-darkening coefficients in these tables cannot be used directly. Claret defines the limb to occur at $\mu^{\prime} = \mu_{\rm cri}$ and set $I_{\lambda}(\mu^{\prime}) = 0 $ for $\mu^{\prime} > \mu_{\rm cri}$. To calculate $h^{\prime}_1$ and $h^{\prime}_2$ the independent variable must be re-scaled using
\[ \mu = (\mu^{\prime}-\mu_{\rm cri})/(1-\mu_{\rm cri}), \]
so $\mu = \nicefrac{2}{3}$ corresponds to $\mu^{\prime} = \nicefrac{2}{3}+\nicefrac{1}{3}\mu_{\rm cri}$ and  $\mu = \nicefrac{1}{3}$ corresponds to  $\mu^{\prime} = \nicefrac{1}{3} +  \nicefrac{2}{3}\mu_{\rm cri}$. This is discussed further in \citet{2023MNRAS.519.3723M}.
The limb-darkening parameters given in \citet{2018A&A...618A..20C} are only for solar metallicity so the following linear correction from \citet{2023MNRAS.519.3723M} was applied to account for the small dependence of limb-darkening on metallicity.
\begin{equation}
    \begin{array}{ll}
        h^{\prime}_{1,{\rm corr}} = h^{\prime}_{1,{\rm cal}} - 0.0027\times[\text{Fe/H}]/0.23 \\
        h^{\prime}_{2,{\rm corr}} = h^{\prime}_{2,{\rm cal}} + 0.0035\times[\text{Fe/H}]/0.23
        \label{eqn:phoenixcorrectiontess}
    \end{array}
\end{equation}
For the models in \citet{2022A&A...666A..60K}, the grids instead give $I_{\lambda}(\mu)$, hence $h^{\prime}_1$ and $h^{\prime}_2$ were computed directly using a linear interpolation. \citet{2022A&A...666A..60K} provide two sets of limb-darkening profiles, ``Set 1'' with a fixed value of the mixing-length parameter and chemical abundances relative to the solar composition from \citet{1998SSRv...85..161G}, and ``Set 2'' using a variable mixing-length parameter depending on T$_{\rm eff}$ \citep[as in,][]{2018ApJ...858...28V} and the solar composition from \citet{2009ARA&A..47..481A}.

The sample of stars targets used in this section to test the accuracy of limb-darkening estimates from models are a combination of the 17 targets in Table~\ref{tab:params} with errors on  $h^{\prime}_1 \le 0.03$ plus 10 systems analysed in the {\it TESS} band by \citet{2023MNRAS.519.3723M}. 
Systems with errors on $h^{\prime}_1>0.03$ (EBLM~J0500$-$46 and EBLM~J0709$-$52) were excluded this analysis of the limb-darkening due to insufficient constraints on the observed limb-darkening profile.
The number of these targets with $T_{\rm eff}$, $\log g$ and $[\text{Fe/H}]$ within the limits of each model grid ($N$) is given in Table~\ref{tab:results_offsets}.

\subsection{Results} \label{subsec:ld_results}
The offset $\Delta h^{\prime}_1 = h^{\prime}_{1,\mathrm{obs}} - h^{\prime}_{1,\mathrm{cal}}$ is defined here as the ``observed'' value of $h^{\prime}_1$ from the analysis of the {\it TESS} light curve minus the ``calculated'' value $h^{\prime}_1$ predicted from a grid of stellar model atmospheres, and similarly for $\Delta h^{\prime}_2$. Summary statistics from the methods in Section~\ref{sec:combine} are given in Table~\ref{tab:results_offsets}. On the whole, the average offsets of $\Delta h^{\prime}_1$ and $\Delta h^{\prime}_2$ found here agree with those from \citet{2023MNRAS.519.3723M} with the average offset uncertainties overlapping. Computing the mean average offsets between models in this study, the offset in $\Delta h_1^{\prime} \approx +0.0086$ and $\Delta h_2^{\prime} \approx -0.0076$. The results in Table~\ref{tab:results_offsets} are similar across all models with $\Delta h^{\prime}_1$ and $\Delta h^{\prime}_2$ within 3 standard deviations of the overall averages just calculated.

Our $\langle \Delta h^{\prime}_1 \rangle$ and $\langle \Delta h^{\prime}_2 \rangle$ compared with \citet{2023MNRAS.519.3723M} generally agree for each model, with the errors on the average offsets overlapping. 

\begin{table*}
\centering
\caption{Results from our light curve analysis combined with {\it TESS} systems from \citet{2023MNRAS.519.3723M}. Targets with standard error $>0.03$ on $ h^{\prime}_1$ were excluded from analysis.}
	\label{tab:results_offsets}
\begin{tabular}{@{}lllrrrrl} 
\hline		
  \multicolumn{1}{@{}l}{Model} & 
  \multicolumn{1}{l}{Source} & 
  \multicolumn{1}{c}{$\langle \Delta h^{\prime}_1\rangle$ } & 
  \multicolumn{1}{c}{$\sigma_{\rm ext,1}$ } & 
  \multicolumn{1}{c}{$\langle \Delta h^{\prime}_2\rangle$ } & 
  \multicolumn{1}{c}{$\sigma_{\rm ext,2}$ } & 
  \multicolumn{1}{c}{$N$ } 
  & Notes
   \\ 
\hline
\noalign{\smallskip}
ATLAS       	    & \text{\citet{2017A&A...600A..30C}}	& $+0.008 \pm 0.002$	& $0.007$	& $-0.013 \pm 0.003$	& $0.001$	& $27$	& Microturbulence $\xi = 2$\,km/s\\
\text{MPS-ATLAS}   	& \text{\citet{2022A&A...666A..60K}}	& $+0.006 \pm 0.002$	& $0.006$	& $-0.010 \pm 0.003$	& $0.001$	& $27$	& Set 1\\
\text{MPS-ATLAS}   	& \text{\citet{2022A&A...666A..60K}}	& $+0.009 \pm 0.002$	& $0.007$	& $-0.011 \pm 0.003$	& $0.001$	& $27$	& Set 2\\
\text{PHOENIX-COND}	& \text{\citet{2018A&A...618A..20C}}	& $+0.014 \pm 0.002$	& $0.007$	& $-0.004 \pm 0.003$	& $0.003$	& $27$	& Linear correction for $[\text{Fe/H}]$\\
\text{Stagger-grid}	& \text{\citet{2018A&A...616A..39M}}	& $+0.006 \pm 0.002$	& $0.007$	& $+0.000 \pm 0.004$	& $0.004$	& $19$	& \\
\noalign{\smallskip}\hline
\end{tabular}
\end{table*}

\begin{figure*}
	\includegraphics[width=\textwidth]{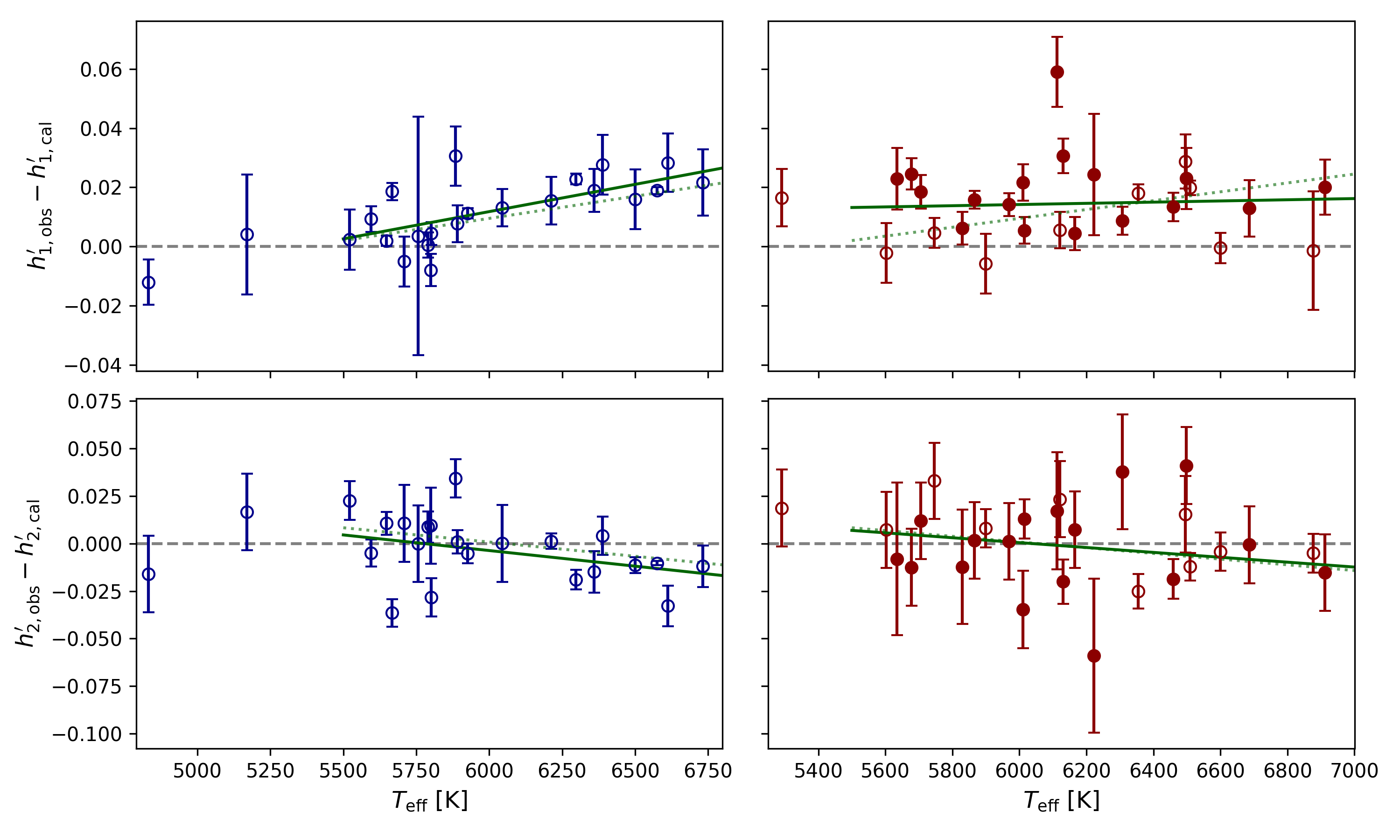}
    \caption{Observed limb-darkening profile compared to the parameters predicted by the \text{PHOENIX-COND} model calculated in \citet{2018A&A...618A..20C}. \textit{Kepler} systems are in blue on the left panels and \textit{TESS} systems are in red on the right. Data from \citet{2023MNRAS.519.3723M} uses unfilled circles, targets characterised in this paper use filled circles. The solid green line is the best fit for each set of systems for $T_{\rm eff}>5500\,$K and the dotted green line is the relation presented in \citet{2023MNRAS.519.3723M}. The grey dashed line is the zero line.} 
    \label{fig:phoenixtrend}
\end{figure*}

As is discussed in \citet{2022A&A...666A..60K} and \citet{2023MNRAS.519.3723M} these offsets can be explained by the impact of the magnetic field on the temperature structure of the star's atmosphere. This opens up the possibility to use limb-darkening to estimate the mean magnetic field strength in the stellar photosphere \citep{2023A&A...679A..65L, 2024NatAs.tmp...73K}.
For stars with higher levels of magnetic activity than those studied here, a better understanding of how star spots and faculae affect the measurement of the limb-darkening profile will be needed before we can interpret these measurements in terms of the star's magnetic field strength \citep{2013A&A...549A...9C}.

\subsubsection{Trends with effective temperature}\label{subsubsec:ld_teff}
In \citet{2023MNRAS.519.3723M}, the relationships between $\Delta h^{\prime}_1$ and $\Delta h^{\prime}_2$, and $T_\text{eff}$ were studied using the combined sample of measurements from {\it Kepler} and {\it TESS} light curves because there were too few stars observed with {\it TESS} in that sample to warrant a separate analysis. 
The linear relation measured for the \text{PHOENIX-COND} model \citep{2018A&A...618A..20C} can be seen in Fig.~\ref{fig:phoenixtrend} as a dotted green line. 
With the addition of 17 of the targets from this study, we are able to investigate the offsets in the {\it TESS} and {\it Kepler} samples separately. 
{\it Kepler} and {\it TESS} have different wavelength response functions so we expect the limb-darkening profiles observed by these two instruments be different. We did not increase the lower range of effective temperature or the sample with $T_\text{eff}<5500K$, so the following fits exclude targets below this effective temperature.

Fig.~\ref{fig:phoenixtrend} shows the comparison between measured and theoretical limb-darkening parameters for the \text{PHOENIX-COND} atmospheric model \citep{2018A&A...618A..20C}.
Looking at Fig.~\ref{fig:phoenixtrend}, we can see a clear trend in $\Delta h^{\prime}_1$ and $\Delta h^{\prime}_2$ for the {\it Kepler} band systems, however the {\it TESS} systems seem to only show an offset with scatter. We fit straight lines to the data to determine if the trends with effective temperature are significant. We use a weighted least squares linear fit adding scatter, $\sigma_{1}$ and $\sigma_{2}$, to the standard error on $\Delta h^{\prime}_1$ and $\Delta h^{\prime}_2$ by quadrature to achieve a fit with $\chi^2 = N_{\rm df}$. For the {\it Kepler} systems we find
\begin{equation}
    \begin{array}{c}
    \label{eqn:kepler_phoenix}
        \Delta h^{\prime}_{1} = (+0.0118 \pm 0.0017) + (+0.0184 \pm 0.0048) Y \\
		\Delta h^{\prime}_{2} = (-0.0037 \pm 0.0037) + (-0.0164 \pm 0.0097) Y
    \end{array}
\end{equation}
where $Y = (T_{\rm eff}-6000{\rm K})/1000{\rm K}$, with $\sigma_1 = 0.00524$ and $\sigma_2 = 0.01359$. Hence, there is potentially a significant relationship ($>3$ standard deviations from the gradient encompassing zero) between $\Delta h^{\prime}_{1}$ and $T_{\rm eff}$ for the Kepler targets and a not significant relation between $\Delta h^{\prime}_{2}$ and $T_{\rm eff}$. For the {\it TESS} systems we find
\begin{equation}
    \begin{array}{c}
    \label{eqn:tess_phoenix}
        \Delta h^{\prime}_{1} = (+0.014 \pm 0.003) + (+0.002 \pm 0.007) \,Y, \\
		\Delta h^{\prime}_{2} = (+0.001 \pm 0.004) + (-0.013 \pm 0.010) \,Y,
    \end{array}
\end{equation}
with $\sigma_1 = 0.0097$ and $\sigma_2 = 0.008$, i.e. there seems to be no significant trend in limb-darkening parameter discrepancy with effective temperature in the {\it TESS} band.

A series of many overlapping absorption features (``line blanketing'') can be seen when looking at spectra of F and G dwarf stars at around $400-650\,\text{nm}$. This area of the spectrum is outside the {\it TESS} band \citep[$600-1000\,\text{nm}$;][]{2014SPIE.9143E..20R} whereas {\it Kepler} observed at shorter wavelengths \citep[$400-900\,\text{nm}$;][]{2008IAUS..249...17B} which include this wavelength region. For this reason we speculate that inaccurate characterisation of the line blanketing in the PHOENIX-COND, model may be the reason for why we see a trend with $T_\text{eff}$ in the {\it Kepler} band but not the {\it TESS} band.

Trends between $T_{\rm eff}$ and $\Delta h^{\prime}_1$ or $\Delta h^{\prime}_2$ for all the other models are are omitted as they are likely insignificant as their gradients encompass zero within 1.5 standard deviations.

\subsubsection{Trends with metallicity}
Using a similar method to Section~\ref{subsubsec:ld_teff}, we fit linear polynomials to check trends of $\Delta h^{\prime}_1$ and $\Delta h^{\prime}_2$ with metallicity. We find there is likely no significant trends for all systems investigated in the {\it TESS} band, where a target is within the confines of a model grid (the number of targets can be seen in column $N$ of Table~\ref{tab:results_offsets}). All gradients encompass zero gradient within one standard deviation, except for the \text{PHOENIX-COND} atmospheric model \citep{2018A&A...618A..20C} which instead requires $1.1$ standard deviations. {\it Kepler} targets were not investigated in this section as no targets were added to increase the metallicity sample.

\section{Radius inflation} \label{sec:rad_infl}
We estimate parameters for the M-dwarf secondary star in our EBLM systems to investigate the radius inflation problem. For this analysis, the targets were re-fit using the same methods as in Section~\ref{subsec:fit} but assuming that limb-darkening is accurately modelled by the power-2 law:
\begin{equation}
\label{eqn:pw2}
    I_{\lambda}(\mu) = 1-c\left(1-\mu^{\alpha}\right).
\end{equation}

By making this assumption, we can obtain more precise estimates of the stellar radii than if we use the much weaker assumptions used above for our investigation of limb-darkening.
The power-2 law is chosen specifically as it outperforms other 2-parameter laws across visible and infrared wavelength ranges for F-type stars and cooler, like the host stars in our sample  \citep{2017AJ....154..111M}.

The limb-darkening parameters  $\alpha$ and $c$ in equation \ref{eqn:pw2} are highly correlated in the fits to a typical transit light curve \citep{2018A&A...616A..39M} so we instead use $q_1$ and $q_2$ defined as follows to parameterise the limb-darkening,  which are much less correlated:  
\begin{equation}
\label{eqn:q1q2}
\begin{array}{l}
     \bmath{q_1 = (1-c2^{-\alpha})^2,} \\
     \noalign{\smallskip}
     \bmath{q_2 = \frac{1-c(1-2^{-\alpha})-c2^{-\alpha}}{1-c2^{-\alpha}}.}
\end{array}
\end{equation}

These parameters also have the advantage that the valid range for both parameters is $0<q_{1,2}<1$ \citep{2019RNAAS...3..117S}.

Refitting with the power-2 law required more steps for the {\sc emcee} chains to achieve convergence than fitting with the previous law. Generally we used 50,000 steps in the {\sc emcee} chains, but we used more steps if needed in cases where this was not enough for the chains to have converged, up to a maximum of 100,000 steps in the case of EBLM~J0723$+$79.

\subsection{Secondary star effective temperature estimates} \label{subsec:Teff2}
The surface brightness ratio, $J=L/k^2$, is closely related to the ratio of the stars' effective temperatures $T_{\rm eff,2}/T_{\rm eff,1}$, particularly for measurements of $J$ in the Rayleigh-Jeans tail of the stars' spectral energy distributions, as is the case of the {\it TESS} band.
We use synthetic photometry from the BT-NextGen (AGSS2009) model grid\footnote{Accessed through the Spanish Virtual Observatory (SVO)} to estimate $T_{\rm eff,2}$ based on our estimates of $T_{\rm eff, 1}$ and the surface brightness ratio in the {\it TESS} band $J_T = S_{T,2}/S_{T,1}$.  
We chose the BT-NextGen (AGSS2009) grid of models because they cover a wide range of effective temperature, surface gravity, metallicity and $\alpha$-element abundance. These models are further described in \citet{2011ASPC..448...91A,2012RSPTA.370.2765A}.

\begin{figure*}
        \includegraphics[width=\textwidth]{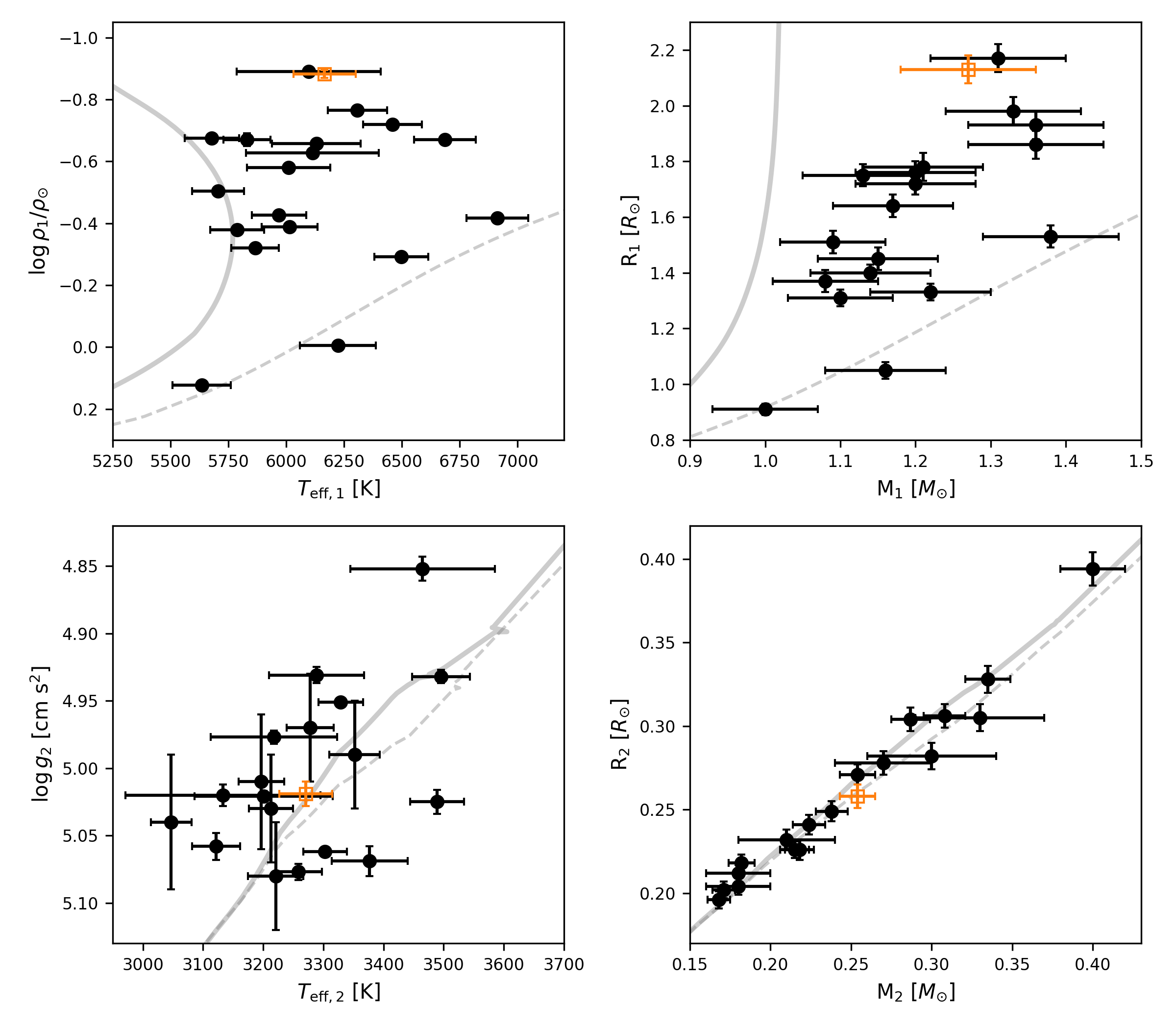}
    \caption{Parameters determined for the primary and secondary stars, plotted with isochrones for comparison. Black filled circle points with error bars are EBLMs analysed using synthetic photometry assuming solar $\alpha$-element abundance ([$\alpha$/Fe] = 0.0). The orange unfilled square points and error bars show target EBLM~J0432$-$33 using synthetic photometry at [$\alpha$/Fe] = $+0.2$ due to evidence of an enhanced $\alpha$-element abundance in the literature. The grey lines are MIST isochrones \citep{2016ApJS..222....8D,2016ApJ...823..102C} plotted for ages 10Gyr (solid line) and 1Gyr (dashed line) at solar metallicity. The top panels give parameters of the primary star. The bottom panels give parameters of the secondary star.}
    \label{fig:massradius}
\end{figure*}

The linear interpolator {\sc LinearNDInterpolator} from {\sc scipy}\footnote{https://scipy.org/} was used to build a grid of surface brightness in the TESS band, $S_T$, as a function of $T_{\rm eff}$, $[\text{Fe/H}]$, $\log g$ and $[\alpha\text{/Fe}]$. 
$[\alpha\text{/Fe}]$ was available in increments of 0.2 in the range $-0.2$ to $+0.6$. Unless a target showed evidence of a non-solar relative $\alpha$-element abundance in the literature, we assume targets have $[\alpha\text{/Fe}] = +0$. The exception was EBLM~J0432$-$33 shown with an orange unfilled square in Fig.~\ref{fig:massradius} for which there is evidence of an enhanced relative $\alpha$-element abundance (+0.24 in \citet{2024MNRAS.531.4085F}). For EBLM~J0432$-$33 we used the $[\alpha\text{/Fe}] = +0.2$ model.

Some of our systems were outside of the model grid range, having too high of a metallicity. We used a least squares fit to a first degree polynomial to extrapolate the metallicity at $[\alpha\text{/Fe}] = +0$. This extrapolated grid was necessary to determine $T_{\rm eff,2}$ for EBLM~J0247$-$51, EBLM~J0440$-$48 and EBLM~J0941$-$51 as noted in results Table~\ref{tab:massradius}. 

The surface brightness was extracted from the grid for a target's $T_{{\rm eff},1}$, $[\text{Fe/H}]$ and $\log g_1$. The error was obtained by first finding $S_{T,1}$ when the input values for $[\text{Fe/H}]$ and $\log g_1$ were kept at their best estimate but $T_{{\rm eff},1}$ with the error added (or subtracted if $T_{{\rm eff},1} + \sigma_{T_{{\rm eff},1}}$ was outside the grid), giving the error in $S_{T,1}$ contributed by $T_{{\rm eff},1}$. 
This was repeated for $[\text{Fe/H}]$ and $\log g_1$, then the three uncertainties added in quadrature to give $\sigma_ {S_{T,1}}$. 
Multiplying $S_{T,_1}$ by $J_T$ found in Section~\ref{sec:rad_infl} gives $S_{T,2}$ with the fractional errors added in quadrature to get the error, $\sigma_{S_{T,2}}$.
We then used a root finding procedure to find $T_{{\rm eff},2}$ from the $S_T(T_{\rm eff}$, $[\text{Fe/H}]$, $\log g$ and $[\alpha\text{/Fe}])$ grid where $[\text{Fe/H}]$, $\log g$ and $[\alpha\text{/Fe}]$ are constants for the secondary star. For this we assumed metallicity and $\alpha$-element abundance were the same as for the primary star and use the value of $\log g_2$ found in Section~\ref{subsec:MRg}.
A similar method to find the error on $T_{{\rm eff},2}$ was used as for $\sigma_{S_{T,2}}$. The values of $T_{{\rm eff},2}$
estimated with this method are given in Table~\ref{tab:massradius} and are shown in Fig.~\ref{fig:massradius}.

\subsubsection{Assessing the impact of tidal distortion} \label{subsubsection:tidaldistortion}
The  {\sc batman} light curve model assumes that both the primary star and its companion are spherical bodies. 
This is a very good approximation for slowly-rotating stars with planet-mass companions but will be less accurate for stars with more massive companions in short-period systems because the two stars will be distorted by their mutual gravitational interaction. 
We have quantified and corrected for the impact of this approximation on our results by analysing simulated light curves that include the effects of tidal distortion and analysing these in the same way as the observed light curves. 
To simulate the light curves we use the {\tt ellc} binary star model \citep{2016A&A...591A.111M} that uses biaxial ellipsoids to approximate the shape of the stars. 
We used the option to compute the shape of the stars assuming that the stars are polytropes with a polytropic index $n=1.5$.
Gravity darkening is accounted for assuming that the specific intensity is related to the local surface gravity by a power law with exponent $y$.
The value of $y$ for the primary star is taken from \citet{2011A&A...529A..75C} using the $I$ band as an approximation for the {\it TESS} response function. 
Similarly, we used the $I$-band limb-darkening coefficients from the same source to model limb-darkening on the primary star using Claret's 4-parameter law.
The limb-darkening and gravity-darkening effects on the M-dwarf companion have a negligible effect on the light curve and so were ignored. 
Tidal distortion increases rapidly as a function of $R_1/a$ so we only generated simulated light curves for systems with $R_1/a > 0.14$.
We used the parameters from Table~\ref{tab:fitresults} to generate a model light curve for these binary systems with the same time sampling as the points around the eclipses that we analysed for the observed light curve.
We then used the residuals from the best-fit light curve to account for the noise in the observed light curve. These simulated light curves were then analysed in the same way as the observed light curves. 
For each system, we also simulated and analysed a light curve in the same way ignoring tidal distortion and gravity darkening. 
We could then use the difference in the results obtained from these two simulated light curves to quantify and correct for the impact of tidal distortion on our results.
Corrections are presented in column $\Delta$\,T$_{\rm eff,2}$ of Table \ref{tab:massradius}.
We find that this temperature correction  becomes insignificant in comparison to the error on $T_{\rm eff,2}$ for systems with $R_1/a < 0.14$. 
The difference between the values obtained from the fit to simulated light curves using distorted and spherical models for the parameters $R_1/a$, $h^{\prime}_1$ and $h^{\prime}_2$ are much less than their standard errors.

\subsection{Mass and radius estimates} \label{subsec:MRg}
The values of $R_1/a$, $R_2/a$ and $i$ from the analysis of the light curve plus the semi-amplitude of the primary star's spectroscopic orbit ($K_1$) can be used to estimate the mass and radius of both stars if we have a reliable estimate for either the mass or radius of the primary star. 
For our analysis, we use an estimate for the primary star mass, $M_1$, based on an empirical relation between the mean stellar density, T$_{\rm eff}$ and metallicity for solar-type stars. 
The mean stellar density of the primary star can be determined directly from $R_1/a$ and  the mass ratio, $q = M_2/M_1$. The dependence on $q$ is weak for small value of $q$, so values of $M_1$, $R_1$, $M_2$ and $R_2$ can be found that are consistent with all the observational constraints given any reasonable initial estimate for $M_1$ and a single iteration of the following procedure.

From the fit in Section~\ref{sec:rad_infl}, $k$, $R_1/a$, $e$, $J$, $\sin{i}$ and their relevant errors and $P$ were extracted as in Section~\ref{subsec:fit}. 
The following parameters and errors were also collected: $T_{\rm eff,1}$, $\log g_1$ and $[\text{Fe/H]}$ from Table~\ref{tab:alixparam}; $K_1$ from \citet{2017A&A...608A.129T} or the \citet{2022yCat.1357....0G}.
We found discrepancies between the values of $K_1$ in \citet{2017A&A...608A.129T} and \citet{2022yCat.1357....0G}, with only three of the eight targets  in common between these data sets having $K_1$ consistent within the quoted standard errors at the 1-$\sigma$ level. 
The offsets appear to be random so we treat this discrepancy as an additional noise source for the  Gaia $K_1$ measurements. Across the eight targets, adding noise in quadrature equal to $1.7\,{\rm km\,s}^{-1}$ to the Gaia $K_1$ measurements results in reduced $\chi^2$ value $\chi^2_r = 1$ for the differences between the two data sets. This additional noise of $1.7\,{\rm km\,s}^{-1}$ is then used to inflate the standard error estimate on  the values of $K_1$ that we have used from \citet{2022yCat.1357....0G}.
The radius of the primary star, $R_1$, was estimated from $M_1$ and $\rho_{_1}$. 

The {\sc pycheops} function {\tt massradius} is used for all these calculations because this makes it easy to use a Monte Carlo approach to propagate errors. For each parameter with an error fed into the function, a Gaussian sample of parameters with mean equal to the parameter value and standard deviation equal to the standard error are generated for 100,000 trial values. If no error is given on a parameter (such as $P$ in our case), calculations all use the same value of the parameter.

We feed the sample of trial values into {\tt massradius} with a nominal initial estimate of the primary star mass, $M_1$, to generate a sample of initial estimates for the primary star mean densities, $\langle \rho_1 \rangle$:
\begin{equation}
    \label{eqn:meandensity}
    \langle \rho_1 \rangle = \frac{3 \pi }{G P^2 (1+q) (R_1/a)^3},
\end{equation}
where $G$ is the gravitational constant and $q=M_2/M_1$ is the mass ratio. We then use the empirical relation by \citet{2010A&A...516A..33E} to make an improved estimate of $M_1$ from the values of $\langle \rho_1 \rangle$, [Fe/H] and T$_{\rm eff, 1}$. This calculation is done for all trial values of $\langle \rho_1 \rangle$ using a sample of trail [Fe/H] and T$_{\rm eff, 1}$ values sampled randomly from independent Gaussian distributions. We then add scatter to these trial $M_1$ values sampled from a Gaussian distribution for $\log (M_{\star}/M_{\odot})$ with a mean of 0 and a standard deviation of 0.023 to account for the observed scatter around the empirical relation from \citet{2010A&A...516A..33E}. 
Re-running the {\tt massradius} function with these improved estimates for $M_1$, we obtain the mean and standard deviations given by the function for $M_1$, $R_1$, $M_2$, $R_2$ and $\log g_2$, summarised in Table~\ref{tab:massradius}. We also extract the samples of $M_2$ and $R_2$ from {\tt massradius} for use in Section~\ref{subsec:mist_comp}.

\begin{table*}
\centering
\caption{Data behind Fig.~\ref{fig:massradius}. EBLM~J0432$-$33 used $[\alpha\text{/Fe}] = +0.2$ synthetic photometry to determine temperature $T_{\rm eff,2}$ in line with literature values while the remaining targets assumed $[\alpha\text{/Fe}] = +0.0$. $\Delta T_{\rm eff,2}$ gives the correction that has been applied to column $T_{\rm eff,2}$ based on Section~\ref{subsubsection:tidaldistortion}
differences in results between {\sc batman} fit of spherical model and tidally distorted model based on the parameters.}
 \label{tab:massradius}
 \addtolength\tabcolsep{1pt}
\begin{tabular}{@{}lrrrrrrrrrrr}
\hline		
EBLM ID & 
  \multicolumn{1}{r}{$M_1$ [${\rm M}_{\sun}$]} & 
  \multicolumn{1}{r}{$R_1$ [${\rm R}_{\sun}$]} &
  \multicolumn{1}{r}{$\log \left( \frac{\rho_{_1}}{\rho_{_{\odot}}} \right) $} & 
  \multicolumn{1}{r}{$M_2$ [${\rm M}_{\sun}$]} &
  \multicolumn{1}{r}{$R_2$ [${\rm R}_{\sun}$]} &
  \multicolumn{1}{r}{$\log g_2$} &
  \multicolumn{1}{r}{$T_{\rm eff,2}$ [K]} &
  \multicolumn{1}{r}{$\Delta T_{\rm eff,2}$ [K]} \\ 
		\hline
J0228$+$05	&1.38(9)	& 1.53(4)	& $-$0.417(9)	& 0.168(7)	&0.196(5)	&5.077(6)	&3259(39)	&  $-$  	\\
J0247$-$51	&1.20(8)	& 1.76(4)	& $-$0.658(6)	& 0.224(10)	&0.241(6)	&5.021(4)	&3201(115)\makebox[0pt][l]{$^{\rm a}$}	&$-$23\\ 
J0400$-$51	&1.17(8)	& 1.64(4)	& $-$0.579(6)	& 0.254(11)	&0.271(6)	&4.977(5)	&3218(105)	&$-$27\\
J0432$-$33	&1.27(9)	& 2.13(5)	& $-$0.882(13)	& 0.254(11)	&0.258(7)	&5.019(9)	&3271(44)	&  $-$  	\\
J0440$-$48	&1.20(8)	& 1.72(4)	& $-$0.628(11)	& 0.182(8)	&0.218(5)	&5.020(8)	&3133(162)\makebox[0pt][l]{$^{\rm a}$}	&$-$86\\ 
J0500$-$46	&1.08(7)	& 1.37(4)	& $-$0.379(14)	& 0.171(7)	&0.202(5)	&5.058(10)	&3122(40)	&  $-$  	\\
J0526$-$34	&1.33(9)	& 1.98(5)	& $-$0.765(7)	& 0.335(14)	&0.328(8)	&4.932(5)	&3496(48)	&  $-$  	\\
J0608$-$59	&1.10(7)	& 1.31(3)	& $-$0.320(4)	& 0.308(13)	&0.306(7)	&4.951(2)	&3329(37)	&  $-$  	\\
J0625$-$43	&1.13(8)	& 1.75(4)	& $-$0.674(7)	& 0.287(12)	&0.304(7)	&4.931(6)	&3289(79)	&$-$14\\ 
J0627$-$67	&1.22(8)	& 1.33(3)	& $-$0.292(4)	& 0.215(9)	&0.226(5)	&5.062(3)	&3303(36)	&  $-$  	\\
J0709$-$52	&1.31(9)	& 2.17(5)	& $-$0.890(11)	& 0.40(2)	&0.394(10)	&4.852(9)	&3465(120)	&  $-$  	\\
J0723$+$79	&1.14(8)	& 1.40(3)	& $-$0.388(9)	& 0.27(3)	&0.278(7)	&4.97(4)	&3278(39)	&  $-$  	\\
J0829$+$66	&1.00(7)	& 0.91(2)	& $+$0.123(13)	& 0.18(2)	&0.204(5)	&5.08(4)	&3221(46)	&  $-$  	\\
J0941$-$31	&1.36(9)	& 1.86(5)	& $-$0.670(13)	& 0.238(10)	&0.249(6)	&5.025(9)	&3489(45)\makebox[0pt][l]{$^{\rm a}$}	&  $-$  	\\
J0955$-$39	&1.16(8)	& 1.05(3)	& $-$0.005(14)	& 0.218(9)	&0.226(6)	&5.069(11)	&3377(63)	&  $-$  	\\
J1626$+$57	&1.15(8)	& 1.45(4)	& $-$0.426(12)	& 0.18(2)	&0.212(5)	&5.04(5)	&3047(34)	&  $-$  	\\
J1640$+$49	&1.36(9)	& 1.93(5)	& $-$0.719(12)	& 0.21(3)	&0.232(6)	&5.03(4)	&3213(37)	&  $-$  	\\
J1705$+$55	&1.09(7)	& 1.51(4)	& $-$0.503(14)	& 0.33(4)	&0.305(8)	&4.99(4)	&3352(42)	&  $-$  	\\
J1850$+$50	&1.21(8)	& 1.78(5)	& $-$0.67(2)	& 0.30(4)	&0.282(8)	&5.01(5)	&3197(38)	&  $-$  	\\
 \hline
 \noalign{\smallskip}
 \multicolumn{8}{@{}l}{$^{\rm a}$ Extended grid used to find temperature, as outlined in Section~\ref{subsec:Teff2}}\\
 \end{tabular}
\end{table*}

\subsection{Quantifying radius inflation across our sample} \label{subsec:mist_comp}
\begin{figure}
    \includegraphics[width=\columnwidth]{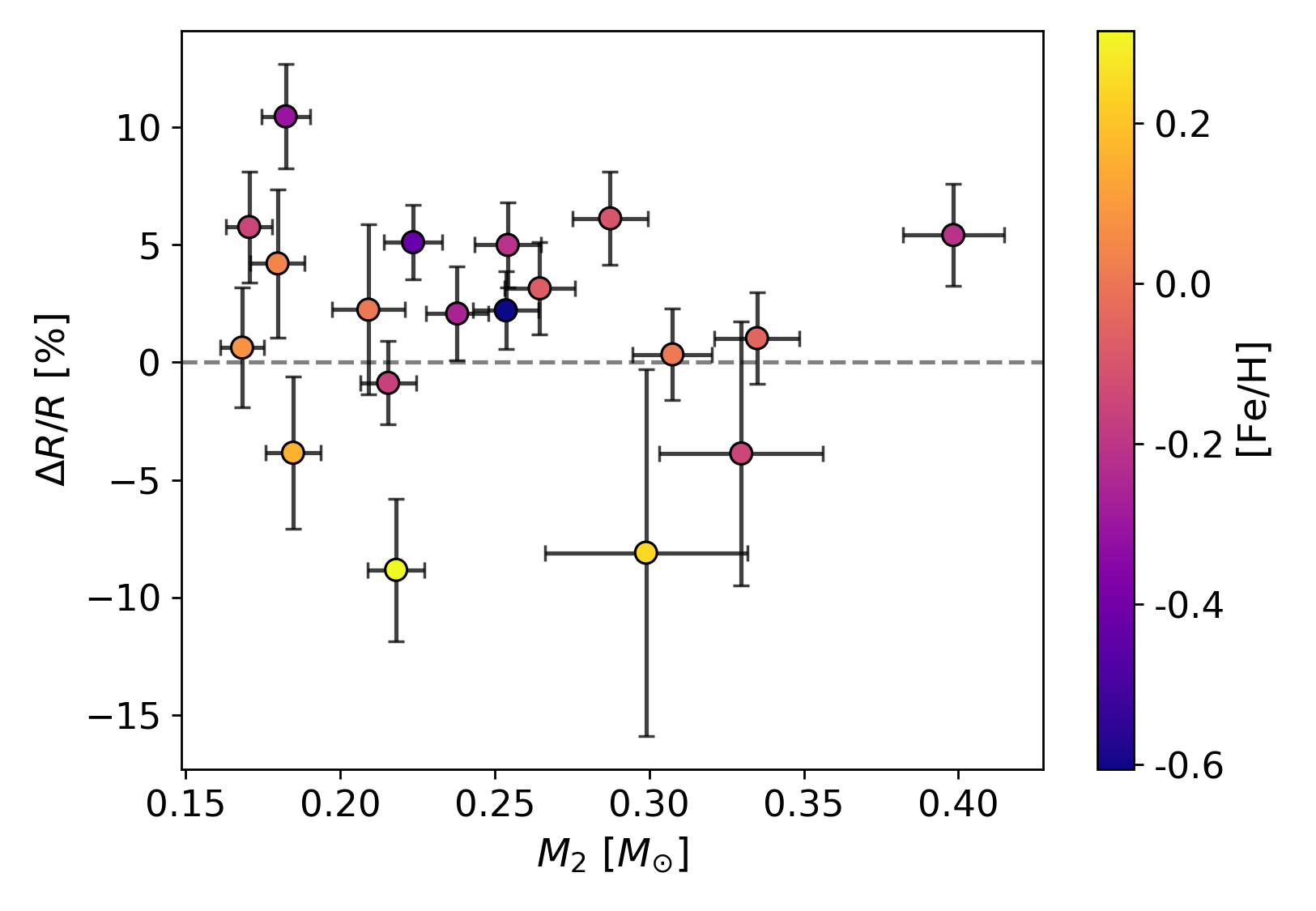}
    \caption{Plot of the radius inflation ($\Delta R/R[\%]$) vs mass ($M_2$) for the secondary star in the full sample of 19 EBLMs. Points are plotted with colour mapping correlated to metallicity.} 
    \label{fig:radiusinflation}
\end{figure}

We compare the secondary star masses and radii in Table~\ref{tab:massradius} to those predicted by MIST models \citep{2016ApJS..222....8D,2016ApJ...823..102C}.\footnote{\url{https://waps.cfa.harvard.edu/MIST/}}
This grid of models was selected because it is widely used and has wide coverage in terms of mass and composition. We use the basic package of theoretical isochrones with parameters $[\alpha\text{/Fe}] = 0$ and initial rotation $v/v_{\rm crit} = 0$.
To ensure we do not introduce a trend with metallicity by comparing masses and radii with $[\text{Fe/H}]$, theoretical radii are extracted for a particular metallicity using interpolation between theoretical isochrones. To build a grid of MIST $R(M,[\text{Fe/H}])$, we use a metallicity range of $-$1.5 to +0.5 in steps of 0.25 and use the {\tt LinearNDInterpolator} function. 
A wider range of metallicities was required to propagate errors in MIST radii due to the Gaussian samples of parameters taken. 
We extract radii from the generated grid for 100 linearly spaced metallicities and masses in the ranges $0.0 - 0.5$ and $0.13 - 0.54\,{\rm M}_{\sun}$ respectively. 
We then use a linear least squares polynomial fit to extrapolate the metallicity to $+1.5$, which was added to the MIST $R(M,[\text{Fe/H}])$ grid allowing enough space for the tails of the  Gaussian samples.
We feed the $M_2$ and $[\text{Fe/H}]$ samples from Section~\ref{subsec:MRg} through the $R(M,[\text{Fe/H}])$ grid to get a sample of MIST radii. Using the sample of ``observed'' radii ($R_{2,\rm obs}$), we gain a sample of the radius inflation, $\Delta R/R[\%]$ for a target,
\begin{equation}
    \label{eqn:perinflation}
    \Delta R/R[\%] = \frac{R_{2,\rm obs} - R_{2,\rm MIST}}{R_{2,\rm MIST}} \times 100
\end{equation}
From these we calculate the mean and standard deviation to find $\Delta R/R[\%]$ and its error and plotted against $M_2$ in Fig.~\ref{fig:radiusinflation}.

Using the {\tt combine} function, described in Section~\ref{sec:combine}, we sample  $\Delta R/R[\%]$ for 3000 steps, discarding the first 2000. From the mean of the sample we get a radius inflation of $2.6 \pm 1.1\%$ 
 with 95\% of the sample showing a radius inflation $<4.4\%$ 

\subsubsection{Metallicity trends}
Looking at Fig.~\ref{fig:radiusinflation} we can see that for our sample of M-dwarfs, stars with higher metallicity tend to have smaller radii than predicted by the MIST models. 
To check the significance of this trend we fit a straight line using a least squares approach, adding scatter equal to 2.24~per~cent to the standard error on the radius inflation to obtain a reduced chi-squared for the fit $\chi^2_r=1$.
We obtain the radius inflation as a function of metallicity as
\begin{equation}
    \label{eqn:inflation_metal}
    \Delta R/R[\%] = (0.9 \pm 1.1) + (-10.6 \pm 3.8)[\text{Fe/H}]
\end{equation}
which had a gradient 2.8 standard deviations from encompassing zero. While this result is only marginally significant, it agrees with \citet{2024MNRAS.528.5703S} who find a likely statistically significant relationship (gradient of $-8.9 \pm 2.9$, converted to be consistent with our units in equation~\eqref{eqn:inflation_metal}). As we use a very similar method to \citet{2024MNRAS.528.5703S} to determine radii, we cannot rule out their comment that different methods of study to investigate radius inflation in M-dwarf stars result in different outcomes.

\section{Conclusions}
In this study we have used {\it TESS} light curves to investigate model predictions versus observations of limb-darkening in F/G dwarf stars in EBLM systems. We have also used the M-dwarf companion stars in these systems to investigate whether the so-called ``radius inflation problem'' noted for low-mass stars extends to stars near the bottom of the main sequence.

We use the summary parameters $h_1^{\prime} = I_{\lambda}(\nicefrac{2}{3})$ and $h_2^{\prime} = h_1^{\prime} - I_{\lambda}(\nicefrac{1}{3})$ to compare the observed limb-darkening profile with those predicted by atmospheric models. Across the models tested we see small but significant offsets in $\Delta h_1^{\prime}$ of $\sim 0.0086$ and $\Delta h_2^{\prime}$ of $\sim -0.0076$. \citet{2022A&A...666A..60K} Set 1 and \citet{2018A&A...616A..39M} perform particularly well, only having small offsets in both $\Delta h_1^{\prime}$ and $\Delta h_2^{\prime}$. In the {\it TESS} band we find no significant trends with $T_\text{eff}$ or $[\text{Fe/H}]$ for any of the models investigated (ATLAS, \citet{2017A&A...600A..30C}; \text{MPS-ATLAS}, \citet{2022A&A...666A..60K} Set 1 and Set 2; \text{PHOENIX-COND}, \citet{2018A&A...618A..20C}; \text{Stagger-grid}, \citet{2018A&A...616A..39M}). In the {\it Kepler} band however the discrepancy between $h_1^{\prime}$ from observations and those predicted by \citet{2018A&A...618A..20C}, have a potentially significant relationship with $T_\text{eff}$.

We also aimed to increase metallicity range of targets by adding a sample of stars from systems less likely to be metal rich in order to test $\Delta h_1^{\prime}$ and $\Delta h_2^{\prime}$ trends with metallicity. The initial sample of stars in the {\it TESS} band from the study by \citet{2023MNRAS.519.3723M} had a metallicity range of $-0.11$ to +0.44 with an average of +0.16. We extended this range on the lower end to $-0.61$ leading to an overall average of 0.004. We found no significant trend with metallicity for any of the grids of limb-darkening parameters tested.

We characterised the M-dwarf secondary component of the binary to determine if our sample of stars shows evidence of the radius inflation problem. Taking the metallicity of our targets into account when comparing with model isochrones, we find a small but significant lack of agreement between the model isochrones and our observations of $2.6 \pm 1.1\%$. We do not see any obvious trend in $T_\text{eff,2}$, however there is scatter of about 100 K around the values predicted by the MIST isochrones.

\section*{Acknowledgements}

JF was partly supported by Science and Technology facilities Council (STFC) grant number ST/Y509383/1. 

PM was supported by Science and Technology facilities Council (STFC) grant number ST/Y002563/1.

This research received funding from the European Research Council (ERC) under the European Union's Horizon 2020 research and innovation program (grant agreement No. 803193/BEBOP).

This paper includes data obtained through the TESS Guest investigator programs  G06022 (PI Martin), G05024 (PI Martin), G04157 (PI Martin), G03216 (PI Martin) and G022253 (PI Martin).

This paper includes data collected by the TESS mission, which is publicly available from the Mikulski Archive for Space Telescopes (MAST) at the Space Telescope Science Institure (STScI). Funding for the TESS mission is provided by the NASA Explorer Program directorate. STScI is operated by the Association of Universities for Research in Astronomy, Inc., under NASA contract NAS 5–26555. We acknowledge the use of public TESS Alert data from pipelines at the TESS Science Office and at the TESS Science Processing Operations Center.

This research made use of Lightkurve, a Python package for Kepler and TESS data analysis \citep{2018ascl.soft12013L}.

\section*{Data Availability}

The data underlying this article are available in the following repository:  Mikulski Archive for Space Telescopes -- \url{https://archive.stsci.edu/}  (TESS). 
CORALIE spectra are available on request by e-mail to the authors.



\bibliographystyle{mnras}
\bibliography{paper} 

\begin{thebibliography}{}
\makeatletter
\relax
\def\mn@urlcharsother{\let\do\@makeother \do\$\do\&\do\#\do\^\do\_\do\%\do\~}
\def\mn@doi{\begingroup\mn@urlcharsother \@ifnextchar [ {\mn@doi@} {\mn@doi@[]}}
\def\mn@doi@[#1]#2{\def\@tempa{#1}\ifx\@tempa\@empty \href {http://dx.doi.org/#2} {doi:#2}\else \href {http://dx.doi.org/#2} {#1}\fi \endgroup}
\def\mn@eprint#1#2{\mn@eprint@#1:#2::\@nil}
\def\mn@eprint@arXiv#1{\href {http://arxiv.org/abs/#1} {{\tt arXiv:#1}}}
\def\mn@eprint@dblp#1{\href {http://dblp.uni-trier.de/rec/bibtex/#1.xml} {dblp:#1}}
\def\mn@eprint@#1:#2:#3:#4\@nil{\def\@tempa {#1}\def\@tempb {#2}\def\@tempc {#3}\ifx \@tempc \@empty \let \@tempc \@tempb \let \@tempb \@tempa \fi \ifx \@tempb \@empty \def\@tempb {arXiv}\fi \@ifundefined {mn@eprint@\@tempb}{\@tempb:\@tempc}{\expandafter \expandafter \csname mn@eprint@\@tempb\endcsname \expandafter{\@tempc}}}

\bibitem[\protect\citeauthoryear{{Allard}, {Homeier}  \& {Freytag}}{{Allard} et~al.}{2011}]{2011ASPC..448...91A}
{Allard} F.,  {Homeier} D.,   {Freytag} B.,  2011, in {Johns-Krull} C.,  {Browning} M.~K.,   {West} A.~A.,  eds,  Astronomical Society of the Pacific Conference Series Vol. 448, 16th Cambridge Workshop on Cool Stars, Stellar Systems, and the Sun. p.~91 (\mn@eprint {arXiv} {1011.5405}), \mn@doi{10.48550/arXiv.1011.5405}

\bibitem[\protect\citeauthoryear{{Allard}, {Homeier}  \& {Freytag}}{{Allard} et~al.}{2012}]{2012RSPTA.370.2765A}
{Allard} F.,  {Homeier} D.,   {Freytag} B.,  2012, \mn@doi [Philosophical Transactions of the Royal Society of London Series A] {10.1098/rsta.2011.0269}, \href {https://ui.adsabs.harvard.edu/abs/2012RSPTA.370.2765A} {370, 2765}

\bibitem[\protect\citeauthoryear{{Asplund}, {Grevesse}, {Sauval}  \& {Scott}}{{Asplund} et~al.}{2009}]{2009ARA&A..47..481A}
{Asplund} M.,  {Grevesse} N.,  {Sauval} A.~J.,   {Scott} P.,  2009, \mn@doi [\araa] {10.1146/annurev.astro.46.060407.145222}, \href {https://ui.adsabs.harvard.edu/abs/2009ARA&A..47..481A} {47, 481}

\bibitem[\protect\citeauthoryear{{Blanco-Cuaresma}}{{Blanco-Cuaresma}}{2019}]{BlancoCuaresma2019}
{Blanco-Cuaresma} S.,  2019, \mn@doi [\mnras] {10.1093/mnras/stz549}, \href {https://ui.adsabs.harvard.edu/abs/2019MNRAS.486.2075B} {486, 2075}

\bibitem[\protect\citeauthoryear{{Blanco-Cuaresma}, {Soubiran}, {Heiter}  \& {Jofr{\'e}}}{{Blanco-Cuaresma} et~al.}{2014}]{BlancoCuaresma2014}
{Blanco-Cuaresma} S.,  {Soubiran} C.,  {Heiter} U.,   {Jofr{\'e}} P.,  2014, \mn@doi [\aap] {10.1051/0004-6361/201423945}, \href {https://ui.adsabs.harvard.edu/abs/2014A&A...569A.111B} {569, A111}

\bibitem[\protect\citeauthoryear{{Borkovits}, {Rappaport}, {Hajdu}  \& {Sztakovics}}{{Borkovits} et~al.}{2015}]{2015MNRAS.448..946B}
{Borkovits} T.,  {Rappaport} S.,  {Hajdu} T.,   {Sztakovics} J.,  2015, \mn@doi [\mnras] {10.1093/mnras/stv015}, \href {https://ui.adsabs.harvard.edu/abs/2015MNRAS.448..946B} {448, 946}

\bibitem[\protect\citeauthoryear{{Borucki} et~al.,}{{Borucki} et~al.}{2008}]{2008IAUS..249...17B}
{Borucki} W.,  et~al., 2008, in {Sun} Y.-S.,  {Ferraz-Mello} S.,   {Zhou} J.-L.,  eds, ~ Vol. 249, Exoplanets: Detection, Formation and Dynamics. pp 17--24, \mn@doi{10.1017/S174392130801630X}

\bibitem[\protect\citeauthoryear{{Borucki} et~al.,}{{Borucki} et~al.}{2010}]{2010Sci...327..977B}
{Borucki} W.~J.,  et~al., 2010, \mn@doi [Science] {10.1126/science.1185402}, \href {http://adsabs.harvard.edu/abs/2010Sci...327..977B} {327, 977}

\bibitem[\protect\citeauthoryear{{Chabrier}, {Gallardo}  \& {Baraffe}}{{Chabrier} et~al.}{2007}]{2007A&A...472L..17C}
{Chabrier} G.,  {Gallardo} J.,   {Baraffe} I.,  2007, \mn@doi [\aap] {10.1051/0004-6361:20077702}, \href {https://ui.adsabs.harvard.edu/abs/2007A&A...472L..17C} {472, L17}

\bibitem[\protect\citeauthoryear{{Choi}, {Dotter}, {Conroy}, {Cantiello}, {Paxton}  \& {Johnson}}{{Choi} et~al.}{2016}]{2016ApJ...823..102C}
{Choi} J.,  {Dotter} A.,  {Conroy} C.,  {Cantiello} M.,  {Paxton} B.,   {Johnson} B.~D.,  2016, \mn@doi [\apj] {10.3847/0004-637X/823/2/102}, \href {https://ui.adsabs.harvard.edu/abs/2016ApJ...823..102C} {823, 102}

\bibitem[\protect\citeauthoryear{{Claret}}{{Claret}}{2000}]{2000A&A...363.1081C}
{Claret} A.,  2000, \aap, \href {http://adsabs.harvard.edu/abs/2000A%26A...363.1081C} {363, 1081}

\bibitem[\protect\citeauthoryear{{Claret}}{{Claret}}{2017}]{2017A&A...600A..30C}
{Claret} A.,  2017, \mn@doi [\aap] {10.1051/0004-6361/201629705}, \href {https://ui.adsabs.harvard.edu/abs/2017A&A...600A..30C} {600, A30}

\bibitem[\protect\citeauthoryear{{Claret}}{{Claret}}{2018}]{2018A&A...618A..20C}
{Claret} A.,  2018, \mn@doi [\aap] {10.1051/0004-6361/201833060}, \href {https://ui.adsabs.harvard.edu/abs/2018A&A...618A..20C} {618, A20}

\bibitem[\protect\citeauthoryear{{Claret} \& {Bloemen}}{{Claret} \& {Bloemen}}{2011}]{2011A&A...529A..75C}
{Claret} A.,  {Bloemen} S.,  2011, \mn@doi [\aap] {10.1051/0004-6361/201116451}, \href {https://ui.adsabs.harvard.edu/abs/2011A&A...529A..75C} {529, A75}

\bibitem[\protect\citeauthoryear{{Csizmadia}, {Pasternacki}, {Dreyer}, {Cabrera}, {Erikson}  \& {Rauer}}{{Csizmadia} et~al.}{2013}]{2013A&A...549A...9C}
{Csizmadia} S.,  {Pasternacki} T.,  {Dreyer} C.,  {Cabrera} J.,  {Erikson} A.,   {Rauer} H.,  2013, \mn@doi [\aap] {10.1051/0004-6361/201219888}, \href {https://ui.adsabs.harvard.edu/abs/2013A&A...549A...9C} {549, A9}

\bibitem[\protect\citeauthoryear{{Dotter}}{{Dotter}}{2016}]{2016ApJS..222....8D}
{Dotter} A.,  2016, \mn@doi [\apjs] {10.3847/0067-0049/222/1/8}, \href {https://ui.adsabs.harvard.edu/abs/2016ApJS..222....8D} {222, 8}

\bibitem[\protect\citeauthoryear{{Eastman}, {Gaudi}  \& {Agol}}{{Eastman} et~al.}{2013}]{2013PASP..125...83E}
{Eastman} J.,  {Gaudi} B.~S.,   {Agol} E.,  2013, \mn@doi [\pasp] {10.1086/669497}, \href {https://ui.adsabs.harvard.edu/abs/2013PASP..125...83E} {125, 83}

\bibitem[\protect\citeauthoryear{{Enoch}, {Collier Cameron}, {Parley}  \& {Hebb}}{{Enoch} et~al.}{2010}]{2010A&A...516A..33E}
{Enoch} B.,  {Collier Cameron} A.,  {Parley} N.~R.,   {Hebb} L.,  2010, \mn@doi [\aap] {10.1051/0004-6361/201014326}, \href {https://ui.adsabs.harvard.edu/abs/2010A&A...516A..33E} {516, A33}

\bibitem[\protect\citeauthoryear{{Espinoza} \& {Jord{\'a}n}}{{Espinoza} \& {Jord{\'a}n}}{2015}]{2015MNRAS.450.1879E}
{Espinoza} N.,  {Jord{\'a}n} A.,  2015, \mn@doi [\mnras] {10.1093/mnras/stv744}, \href {https://ui.adsabs.harvard.edu/abs/2015MNRAS.450.1879E} {450, 1879}

\bibitem[\protect\citeauthoryear{{Espinoza} \& {Jord{\'a}n}}{{Espinoza} \& {Jord{\'a}n}}{2016}]{2016MNRAS.457.3573E}
{Espinoza} N.,  {Jord{\'a}n} A.,  2016, \mn@doi [\mnras] {10.1093/mnras/stw224}, \href {https://ui.adsabs.harvard.edu/abs/2016MNRAS.457.3573E} {457, 3573}

\bibitem[\protect\citeauthoryear{{Fabrycky}}{{Fabrycky}}{2010}]{2010arXiv1006.3834F}
{Fabrycky} D.~C.,  2010, \mn@doi [arXiv e-prints] {10.48550/arXiv.1006.3834}, \href {https://ui.adsabs.harvard.edu/abs/2010arXiv1006.3834F} {p. arXiv:1006.3834}

\bibitem[\protect\citeauthoryear{{Feiden} \& {Chaboyer}}{{Feiden} \& {Chaboyer}}{2013}]{2013ApJ...779..183F}
{Feiden} G.~A.,  {Chaboyer} B.,  2013, \mn@doi [\apj] {10.1088/0004-637X/779/2/183}, \href {https://ui.adsabs.harvard.edu/abs/2013ApJ...779..183F} {779, 183}

\bibitem[\protect\citeauthoryear{{Feiden} \& {Chaboyer}}{{Feiden} \& {Chaboyer}}{2014}]{2014ApJ...789...53F}
{Feiden} G.~A.,  {Chaboyer} B.,  2014, \mn@doi [\apj] {10.1088/0004-637X/789/1/53}, \href {https://ui.adsabs.harvard.edu/abs/2014ApJ...789...53F} {789, 53}

\bibitem[\protect\citeauthoryear{{Foreman-Mackey}, {Hogg}, {Lang}  \& {Goodman}}{{Foreman-Mackey} et~al.}{2013}]{2013PASP..125..306F}
{Foreman-Mackey} D.,  {Hogg} D.~W.,  {Lang} D.,   {Goodman} J.,  2013, \mn@doi [\pasp] {10.1086/670067}, \href {https://ui.adsabs.harvard.edu/abs/2013PASP..125..306F} {125, 306}

\bibitem[\protect\citeauthoryear{{Freckelton} et~al.,}{{Freckelton} et~al.}{2024}]{2024MNRAS.531.4085F}
{Freckelton} A.~V.,  et~al., 2024, \mn@doi [\mnras] {10.1093/mnras/stae1405}, \href {https://ui.adsabs.harvard.edu/abs/2024MNRAS.531.4085F} {531, 4085}

\bibitem[\protect\citeauthoryear{{Gaia Collaboration}}{{Gaia Collaboration}}{2022}]{2022yCat.1357....0G}
{Gaia Collaboration} 2022, VizieR Online Data Catalog, \href {https://ui.adsabs.harvard.edu/abs/2022yCat.1357....0G} {p. I/357}

\bibitem[\protect\citeauthoryear{{Goodman} \& {Weare}}{{Goodman} \& {Weare}}{2010}]{2010CAMCS...5...65G}
{Goodman} J.,  {Weare} J.,  2010, \mn@doi [Communications in Applied Mathematics and Computational Science] {10.2140/camcos.2010.5.65}, \href {https://ui.adsabs.harvard.edu/abs/2010CAMCS...5...65G} {5, 65}

\bibitem[\protect\citeauthoryear{{Gray} \& {Corbally}}{{Gray} \& {Corbally}}{1994}]{gray1994}
{Gray} R.~O.,  {Corbally} C.~J.,  1994, \mn@doi [\aj] {10.1086/116893}, \href {https://ui.adsabs.harvard.edu/abs/1994AJ....107..742G} {107, 742}

\bibitem[\protect\citeauthoryear{{Grevesse} \& {Sauval}}{{Grevesse} \& {Sauval}}{1998}]{1998SSRv...85..161G}
{Grevesse} N.,  {Sauval} A.~J.,  1998, \mn@doi [\ssr] {10.1023/A:1005161325181}, \href {https://ui.adsabs.harvard.edu/abs/1998SSRv...85..161G} {85, 161}

\bibitem[\protect\citeauthoryear{{Hoxie}}{{Hoxie}}{1970}]{1970ApJ...161.1083H}
{Hoxie} D.~T.,  1970, \mn@doi [\apj] {10.1086/150609}, \href {https://ui.adsabs.harvard.edu/abs/1970ApJ...161.1083H} {161, 1083}

\bibitem[\protect\citeauthoryear{{Knutson}, {Charbonneau}, {Noyes}, {Brown}  \& {Gilliland}}{{Knutson} et~al.}{2007}]{2007ApJ...655..564K}
{Knutson} H.~A.,  {Charbonneau} D.,  {Noyes} R.~W.,  {Brown} T.~M.,   {Gilliland} R.~L.,  2007, \mn@doi [\apj] {10.1086/510111}, \href {https://ui.adsabs.harvard.edu/abs/2007ApJ...655..564K} {655, 564}

\bibitem[\protect\citeauthoryear{{Kostogryz}, {Witzke}, {Shapiro}, {Solanki}, {Maxted}, {Kurucz}  \& {Gizon}}{{Kostogryz} et~al.}{2022}]{2022A&A...666A..60K}
{Kostogryz} N.~M.,  {Witzke} V.,  {Shapiro} A.~I.,  {Solanki} S.~K.,  {Maxted} P.~F.~L.,  {Kurucz} R.~L.,   {Gizon} L.,  2022, \mn@doi [\aap] {10.1051/0004-6361/202243722}, \href {https://ui.adsabs.harvard.edu/abs/2022A&A...666A..60K} {666, A60}

\bibitem[\protect\citeauthoryear{{Kostogryz} et~al.,}{{Kostogryz} et~al.}{2024}]{2024NatAs.tmp...73K}
{Kostogryz} N.~M.,  et~al., 2024, \mn@doi [Nature Astronomy] {10.1038/s41550-024-02252-5}, \href {https://ui.adsabs.harvard.edu/abs/2024NatAs.tmp...73K} {}

\bibitem[\protect\citeauthoryear{{Kreidberg}}{{Kreidberg}}{2015}]{2015PASP..127.1161K}
{Kreidberg} L.,  2015, \mn@doi [\pasp] {10.1086/683602}, \href {https://ui.adsabs.harvard.edu/abs/2015PASP..127.1161K} {127, 1161}

\bibitem[\protect\citeauthoryear{{Kurucz}}{{Kurucz}}{2005}]{kurucz2005}
{Kurucz} R.~L.,  2005, Memorie della Societa Astronomica Italiana Supplementi, \href {https://ui.adsabs.harvard.edu/abs/2005MSAIS...8...14K} {8, 14}

\bibitem[\protect\citeauthoryear{{Lightkurve Collaboration} et~al.,}{{Lightkurve Collaboration} et~al.}{2018}]{2018ascl.soft12013L}
{Lightkurve Collaboration} et~al., 2018, {Lightkurve: Kepler and TESS time series analysis in Python}, Astrophysics Source Code Library (\mn@eprint {ascl} {1812.013})

\bibitem[\protect\citeauthoryear{{Ludwig}, {Steffen}  \& {Freytag}}{{Ludwig} et~al.}{2023}]{2023A&A...679A..65L}
{Ludwig} H.~G.,  {Steffen} M.,   {Freytag} B.,  2023, \mn@doi [\aap] {10.1051/0004-6361/202346783}, \href {https://ui.adsabs.harvard.edu/abs/2023A&A...679A..65L} {679, A65}

\bibitem[\protect\citeauthoryear{{MacDonald} \& {Mullan}}{{MacDonald} \& {Mullan}}{2014}]{2014ApJ...787...70M}
{MacDonald} J.,  {Mullan} D.~J.,  2014, \mn@doi [\apj] {10.1088/0004-637X/787/1/70}, \href {https://ui.adsabs.harvard.edu/abs/2014ApJ...787...70M} {787, 70}

\bibitem[\protect\citeauthoryear{{Martin} et~al.,}{{Martin} et~al.}{2019}]{2019A&A...624A..68M}
{Martin} D.~V.,  et~al., 2019, \mn@doi [\aap] {10.1051/0004-6361/201833669}, \href {https://ui.adsabs.harvard.edu/abs/2019A&A...624A..68M} {624, A68}

\bibitem[\protect\citeauthoryear{{Maxted}}{{Maxted}}{2016}]{2016A&A...591A.111M}
{Maxted} P.~F.~L.,  2016, \mn@doi [\aap] {10.1051/0004-6361/201628579}, \href {https://ui.adsabs.harvard.edu/abs/2016A&A...591A.111M} {591, A111}

\bibitem[\protect\citeauthoryear{{Maxted}}{{Maxted}}{2018}]{2018A&A...616A..39M}
{Maxted} P.~F.~L.,  2018, \mn@doi [\aap] {10.1051/0004-6361/201832944}, \href {https://ui.adsabs.harvard.edu/abs/2018A&A...616A..39M} {616, A39}

\bibitem[\protect\citeauthoryear{{Maxted}}{{Maxted}}{2023}]{2023MNRAS.519.3723M}
{Maxted} P. F.~L.,  2023, \mn@doi [\mnras] {10.1093/mnras/stac3741}, \href {https://ui.adsabs.harvard.edu/abs/2023MNRAS.519.3723M} {519, 3723}

\bibitem[\protect\citeauthoryear{{Maxted}, {Triaud}  \& {Martin}}{{Maxted} et~al.}{2023}]{2023Univ....9..498M}
{Maxted} P. F.~L.,  {Triaud} A. H.~M.~J.,   {Martin} D.~V.,  2023, \mn@doi [Universe] {10.3390/universe9120498}, \href {https://ui.adsabs.harvard.edu/abs/2023Univ....9..498M} {9, 498}

\bibitem[\protect\citeauthoryear{{Morello}, {Tsiaras}, {Howarth}  \& {Homeier}}{{Morello} et~al.}{2017}]{2017AJ....154..111M}
{Morello} G.,  {Tsiaras} A.,  {Howarth} I.~D.,   {Homeier} D.,  2017, \mn@doi [\aj] {10.3847/1538-3881/aa8405}, \href {https://ui.adsabs.harvard.edu/abs/2017AJ....154..111M} {154, 111}

\bibitem[\protect\citeauthoryear{{Mullan} \& {MacDonald}}{{Mullan} \& {MacDonald}}{2001}]{2001ApJ...559..353M}
{Mullan} D.~J.,  {MacDonald} J.,  2001, \mn@doi [\apj] {10.1086/322336}, \href {https://ui.adsabs.harvard.edu/abs/2001ApJ...559..353M} {559, 353}

\bibitem[\protect\citeauthoryear{{M{\"u}ller}, {Huber}, {Czesla}, {Wolter}  \& {Schmitt}}{{M{\"u}ller} et~al.}{2013}]{2013A&A...560A.112M}
{M{\"u}ller} H.~M.,  {Huber} K.~F.,  {Czesla} S.,  {Wolter} U.,   {Schmitt} J.~H.~M.~M.,  2013, \mn@doi [\aap] {10.1051/0004-6361/201322079}, \href {https://ui.adsabs.harvard.edu/abs/2013A&A...560A.112M} {560, A112}

\bibitem[\protect\citeauthoryear{{Patel} \& {Espinoza}}{{Patel} \& {Espinoza}}{2022}]{2022AJ....163..228P}
{Patel} J.~A.,  {Espinoza} N.,  2022, \mn@doi [\aj] {10.3847/1538-3881/ac5f55}, \href {https://ui.adsabs.harvard.edu/abs/2022AJ....163..228P} {163, 228}

\bibitem[\protect\citeauthoryear{{Popper}}{{Popper}}{1997}]{1997AJ....114.1195P}
{Popper} D.~M.,  1997, \mn@doi [\aj] {10.1086/118552}, \href {http://adsabs.harvard.edu/abs/1997AJ....114.1195P} {114, 1195}

\bibitem[\protect\citeauthoryear{{Queloz} et~al.,}{{Queloz} et~al.}{2000}]{Queloz2000}
{Queloz} D.,  et~al., 2000, \aap, \href {https://ui.adsabs.harvard.edu/abs/2000A&A...354...99Q} {354, 99}

\bibitem[\protect\citeauthoryear{{Ricker} et~al.,}{{Ricker} et~al.}{2014}]{2014SPIE.9143E..20R}
{Ricker} G.~R.,  et~al., 2014, in {Oschmann} Jacobus~M. J.,  {Clampin} M.,  {Fazio} G.~G.,   {MacEwen} H.~A.,  eds,  Society of Photo-Optical Instrumentation Engineers (SPIE) Conference Series Vol. 9143, Space Telescopes and Instrumentation 2014: Optical, Infrared, and Millimeter Wave. p. 914320 (\mn@eprint {arXiv} {1406.0151}), \mn@doi{10.1117/12.2063489}

\bibitem[\protect\citeauthoryear{{Ricker} et~al.,}{{Ricker} et~al.}{2015}]{2015JATIS...1a4003R}
{Ricker} G.~R.,  et~al., 2015, \mn@doi [Journal of Astronomical Telescopes, Instruments, and Systems] {10.1117/1.JATIS.1.1.014003}, \href {http://adsabs.harvard.edu/abs/2015JATIS...1a4003R} {1, 014003}

\bibitem[\protect\citeauthoryear{{Short}, {Welsh}, {Orosz}, {Windmiller}  \& {Maxted}}{{Short} et~al.}{2019}]{2019RNAAS...3..117S}
{Short} D.~R.,  {Welsh} W.~F.,  {Orosz} J.~A.,  {Windmiller} G.,   {Maxted} P.~F.~L.,  2019, \mn@doi [Research Notes of the American Astronomical Society] {10.3847/2515-5172/ab3a3e}, \href {https://ui.adsabs.harvard.edu/abs/2019RNAAS...3..117S} {3, 117}

\bibitem[\protect\citeauthoryear{{Spada}, {Demarque}, {Kim}  \& {Sills}}{{Spada} et~al.}{2013}]{2013ApJ...776...87S}
{Spada} F.,  {Demarque} P.,  {Kim} Y.-C.,   {Sills} A.,  2013, \mn@doi [\apj] {10.1088/0004-637X/776/2/87}, \href {http://adsabs.harvard.edu/abs/2013ApJ...776...87S} {776, 87}

\bibitem[\protect\citeauthoryear{{Stassun} et~al.,}{{Stassun} et~al.}{2019}]{2019AJ....158..138S}
{Stassun} K.~G.,  et~al., 2019, \mn@doi [\aj] {10.3847/1538-3881/ab3467}, \href {https://ui.adsabs.harvard.edu/abs/2019AJ....158..138S} {158, 138}

\bibitem[\protect\citeauthoryear{{Swayne} et~al.,}{{Swayne} et~al.}{2024}]{2024MNRAS.528.5703S}
{Swayne} M.~I.,  et~al., 2024, \mn@doi [\mnras] {10.1093/mnras/stad3866}, \href {https://ui.adsabs.harvard.edu/abs/2024MNRAS.528.5703S} {528, 5703}

\bibitem[\protect\citeauthoryear{{Triaud} et~al.,}{{Triaud} et~al.}{2017}]{2017A&A...608A.129T}
{Triaud} A. H.~M.~J.,  et~al., 2017, \mn@doi [\aap] {10.1051/0004-6361/201730993}, \href {https://ui.adsabs.harvard.edu/abs/2017A&A...608A.129T} {608, A129}

\bibitem[\protect\citeauthoryear{{Verma}, {Maxted}, {Singh}, {Ludwig}  \& {Sable}}{{Verma} et~al.}{2024}]{2024MNRAS.534.3893V}
{Verma} K.,  {Maxted} P. F.~L.,  {Singh} A.,  {Ludwig} H.~G.,   {Sable} Y.,  2024, \mn@doi [\mnras] {10.1093/mnras/stae2344}, \href {https://ui.adsabs.harvard.edu/abs/2024MNRAS.534.3893V} {534, 3893}

\bibitem[\protect\citeauthoryear{{Viani}, {Basu}, {Ong J.}, {Bonaca}  \& {Chaplin}}{{Viani} et~al.}{2018}]{2018ApJ...858...28V}
{Viani} L.~S.,  {Basu} S.,  {Ong J.} M.~J.,  {Bonaca} A.,   {Chaplin} W.~J.,  2018, \mn@doi [\apj] {10.3847/1538-4357/aab7eb}, \href {https://ui.adsabs.harvard.edu/abs/2018ApJ...858...28V} {858, 28}

\bibitem[\protect\citeauthoryear{{Winn}}{{Winn}}{2010}]{2010exop.book...55W}
{Winn} J.~N.,  2010, in {Seager} S.,  ed., , Exoplanets.
pp 55--77, \mn@doi{10.48550/arXiv.1001.2010}

\makeatother
\end{thebibliography}








\bsp	
\label{lastpage}
\end{document}